\documentclass[conference]{IEEEtran}

\usepackage{cite}
\usepackage{amsmath,amssymb,amsfonts}
\usepackage{textcomp}
\usepackage{xcolor}
\usepackage{caption}

\usepackage{amsthm}
\usepackage{siunitx} 
\usepackage[ddmmyyyy]{datetime}
\usepackage{lastpage}
\usepackage{fancyhdr}
\usepackage{balance}
\usepackage[abbreviations]{foreign}
\usepackage{pifont}
\usepackage{graphicx}
\usepackage[font=footnotesize]{subfig}
\usepackage{algorithm}
\usepackage{algpseudocode}
\usepackage{tabularx}
\usepackage{booktabs}
\usepackage{makecell}

\DeclareGraphicsExtensions{.pdf,.jpeg,.png}

\newcommand{\protocol}{\textsc{Raptee}\xspace}
\newcommand{\brahms}{\textsc{Brahms}\xspace}

\begin{document}
\title{\protocol: Leveraging trusted execution environments for Byzantine-tolerant peer sampling services}


\IEEEoverridecommandlockouts


\author{
 \IEEEauthorblockN{
     Matthieu Pigaglio\IEEEauthorrefmark{1},
     Joachim Bruneau-Queyreix\IEEEauthorrefmark{2},
     David Bromberg\IEEEauthorrefmark{3}, \\
     Davide Frey\IEEEauthorrefmark{3},
     Etienne Rivière\IEEEauthorrefmark{1},
     Laurent Réveillère\IEEEauthorrefmark{2}
}

\IEEEauthorblockA{\IEEEauthorrefmark{1} ICTEAM, UCLouvain, Belgium}%
 \IEEEauthorblockA{\IEEEauthorrefmark{2} Univ. Bordeaux, CNRS, Bordeaux INP, LaBRI, UMR5800, F-33400 Talence, France}%
 \IEEEauthorblockA{\IEEEauthorrefmark{3} Univ Rennes, CNRS, Inria, IRISA, Rennes, France}%

}

\maketitle


\begin{abstract}
    Peer sampling is a first-class abstraction used in distributed systems for overlay management and information dissemination.
    The goal of peer sampling is to continuously build and refresh a partial and local view of the full membership of a dynamic, large-scale distributed system.
    Malicious nodes under the control of an adversary may aim at being over-represented in the views of correct nodes, increasing their impact on the proper operation of protocols built over peer sampling. 
    State-of-the-art Byzantine resilient peer sampling protocols reduce this bias as long as Byzantines are not overly present.
    This paper studies the benefits brought to the resilience of peer sampling services when considering that a small portion of trusted nodes can run code whose authenticity and integrity can be assessed within a trusted execution environment, and specifically Intel's software guard extensions technology (SGX).
    We present \protocol, a protocol that builds and leverages trusted gossip-based communications to hamper an adversary's ability to increase its system-wide representation in the views of all nodes.
    We apply \protocol to \brahms, the most resilient peer sampling protocol to date.
    Experiments with 10,000 nodes  show that with only 1\% of SGX-capable devices, \protocol can reduce the proportion of Byzantine IDs in the view of honest nodes by up to 17\% when the system contains 10\% of Byzantine nodes.
    In addition, the security guarantees of \protocol hold even in the presence of a powerful attacker attempting to identify trusted nodes and injecting view-poisoned trusted nodes.
    \end{abstract}

\section{Intro}

Peer sampling represents a first-class abstraction for the construction of large-scale distributed systems.
It is notably employed in distributed unstructured overlay management~\cite{voulgarisVICINITY2013,jelasity2009t} and information dissemination~\cite{matos2013lightweight,eugster2003lightweight}.
Typically, nodes possess partial knowledge (also referred to as their \textit{view}) of the global and dynamic system membership.
The goal of the peer-sampling distributed service is to bootstrap and continuously refresh this local view so that it corresponds as much as possible to a \emph{uniform} sample of alive nodes in the system.
The implementation of the peer-sampling service is typically based on gossip protocols leveraging simple, periodic peer-wise exchanges of information.
A plethora of protocols have been published and studied~\cite{voulgarisCYCLON2005,jelasityGossipbased2007,anceaume2013uniform} under considerations of crash faults, churns, performance, ergodicity, and desirable structural properties such as balanced in-degree, low diameter, and ability to quickly remove departed nodes from the views of alive ones.

The resilience of peer-sampling protocols to Byzantine faults is crucial for the security of applications relying upon it.
Indeed, malicious nodes that succeed in becoming over-represented in the views of honest nodes 
can gain control of the upper-layer protocols.
For example, the peer-sampling protocol of Bitcoin was discovered to be exposed to eclipse attacks~\cite{heilmanEclipse2015}, opening the door to multiple types of selfish mining and double-spending attacks at the consensus level.
In state-of-the-art Byzantine-resilient peer-sampling protocols~\cite{bortnikov2009brahms,anceaume2013uniform}, the views of honest nodes can swiftly be poisoned with Byzantine identifiers as the proportion of malicious nodes in the system grows.
With \brahms~\cite{bortnikov2009brahms}, the most Byzantine resilient protocol, the views of honest nodes contain 81\% of Byzantine identifiers (\textit{IDs}) when 18\% of the nodes in the system are malicious.

The advent of trusted execution environments (TEEs) has exposed many desirable properties for distributed systems, notably remote attestation, authentication, integrity, and confidentiality of the execution environment.
This paper shows how the resilience of peer-sampling protocols to Byzantine behaviors can be improved with these TEE properties.
In this paper, we consider Intel's Software Guard Extensions (SGX)~\cite{costanIntel2016} released in 2015, but any TEE technology providing similar code integrity validation properties would be a fit for our approach, e.g., ARM's TrustZone~\cite{pinto2019demystifying}. 


We present \protocol, a novel protocol that integrates trusted communications in the peer-sampling-service operation and interoperates them with \brahms. 
We consider a system composed of honest and Byzantine nodes, along with a small proportion of trusted nodes (\ie SGX-capable devices).
Because trusted nodes cannot act maliciously, \protocol accelerates the dissemination of knowledge they possess while slowing down the action of Byzantine nodes and preventing the identification of trusted nodes by an adversary. 

In \protocol, all nodes execute a modified version of \brahms. 
However, trusted nodes are not bound to strictly execute \brahms when interacting with other trusted nodes, even though they can possess Byzantine IDs in their views.
Contrarily, when communicating with untrusted nodes, additional Byzantine-fault countermeasures are required. 
They consist of adaptively ignoring part of the IDs sent by untrusted nodes. 
This voluntary loss of information enables higher resilience.
Indeed, it reduces the ability of an attacker to pollute the views of trusted nodes by sending them bulks of malicious IDs, which makes trusted nodes act as sources of less biased identifiers than untrusted nodes.

Because trusted nodes can improve the system's Byzantine resilience, they are of utmost interest for malicious nodes in performing targeted attacks and eclipsing them from the system.
It becomes clear that they cannot advertise themselves as trustworthy actors and should remain hidden in the mass.
Accordingly, \protocol leverages the ability of trusted nodes to learn their mutual trusted capacity without revealing it to others.
Ignoring IDs also hampers the dissemination of correct IDs transmitted by honest nodes, which could have a significant impact on system convergence in the time it takes to discover a majority of nodes.
To address this problem, \protocol makes trusted nodes exchange information in a dissemination-efficient way using gossip-based peer-sampling interactions following the framework of Jelasity \emph{et al.}~\cite{jelasityGossipbased2007}.

Our experiments with 10,000 nodes on the Grid~5000~\cite{bolze2006grid} testbed show that with no more than 1\% of SGX-capable devices, \protocol can reduce the proportion of Byzantine IDs in the views of honest nodes by up to 17\% when the ratio of Byzantine nodes in the system is 10\%. 
This resilience gain comes at the expense of a limited overhead of 10\% in terms of the required number of rounds to view stability and 12\% for system discovery.
We assess the security risks of \protocol against two types of attack: trusted-node identification and view-poisoned trusted node injection.
We show that the identification attack is particularly inefficient before the system converges and is impossible once this point is reached.
Considering injecting trusted nodes with polluted views, we show that it has little to no impact on the system's resilience and that trusted nodes can self-heal with \protocol. 

\noindent
\textbf{Outline.}
First, we provide some requisite background knowledge on \brahms and on the general gossip-based peer-sampling framework in Section~\ref{sec:background}.
Then, we provide a high-level overview of \protocol with its models and objectives in Section~\ref{sec:system}.
We describe \protocol and detail how it complements \brahms with trusted communications in Section~\ref{sec:protocol}. 
Section~\ref{sec:exp} exposes the experimental methodology used to evaluate  the performance and resilience of \protocol under different configurations and discusses the results. 
We provide a security assessment of \protocol to trusted-node-identification and corrupted-trusted-node-injection attacks in Section \ref{sec:sec}. We summarize our findings in Section~\ref{sec:discussion} and review the state of the art in Section~\ref{sec:rw} before concluding in Section~\ref{sec:conclusion}.



\section{Background}
\label{sec:background}



We start by presenting background on gossip-based peer sampling and on the \brahms protocol~\cite{bortnikov2009brahms}.

\noindent
\textbf{Gossip-based peer sampling:}
Gossip is a mode of operation for large-scale distributed systems based on repeated, periodic peer-to-peer interactions between pairs of peers~\cite{riviere2011gossip}.
Despite the general simplicity of these local interactions, Gossip protocols exhibit very good global robustness properties (\ie the ability to resist high levels of churn or survive network partitions) thanks to their self-organizing nature~\cite{jelasity2009t,babaoglu2006design}.
These properties often depend, however, on the quality of the random process that determines the participants in these peer-wise interactions.
This selection of communication partners at each period should be as close as possible to a uniform random draw amongst all participating nodes, including newly joined ones and excluding departed or failed ones.
The \emph{peer-sampling service} plays a fundamental role in ensuring this property.

Interestingly, the most efficient way to implement a peer-sampling service for use by other protocols, gossip-based or not, is to design it as a gossip-based protocol itself.
The goal of such a protocol is to equip each node of the system with a view, \eg a sample of other nodes in the system.
The graph formed by the who-knows-whom relations with these views should be as close as possible to a random graph of fixed out degree.
As such, it should have balanced in-degrees (to prevent nodes from being under or over-represented in the views of other nodes), a low diameter, and a low clustering coefficient.
In addition, these properties must be enforced while quickly including newly joined peers in the views of other nodes (so as to match their in-degree, and therefore the probability of being selected as a gossip partner, to that of existing nodes) and quickly removing departed or failed nodes.
Addressing these different requirements simultaneously is, unfortunately, impossible and gossip-based peer-sampling protocols are the result of a compromise between them.

Jelasity \emph{et al.}~\cite{jelasity2009t} proposed a universal framework for the design of gossip-based protocols that encompasses previous designs with emphasis on a specific property (\eg balanced in degree and low clustering for Cyclon~\cite{voulgarisCYCLON2005} or efficient dynamic membership for Newscast~\cite{tolgyesi2009adaptive}).
This framework defines two parameters, $H$ (for Heal) and $S$ (for Shuffling), that drive the exchange of (partial) views of size $c$ between peers, in addition to other parameters such as the policy for partner selection or the policy for the link deprecation in individual nodes' views.
On the basis of the recommendations of Jelasity \emph{et al.}~\cite{jelasity2009t}, we apply this framework with the following criteria.
  (1)~Peers initiate an exchange with the peer that has been for the longest time in their view, on the basis of an age parameter associated with its entries. This effectively enables a round robin \emph{probing} of view neighbors.
  (2)~They exchange half of their view, and the initiator peer inserts a link to itself in the view sent to the gossip partner.
  (3)~The exchange favors the \emph{shuffling} of the links, \ie a link sent from the initiator will be kept only by the partner, and \emph{vice-versa}.



\noindent
\textbf{\brahms:}
This protocol presented by Bortnikov \emph{et al.}~\cite{bortnikov2009brahms} is designed for large and dynamic systems prone to Byzantine failures and sybil attacks. 
With high probability, \brahms prevents an attacker from creating a partition between correct nodes, and allows each node's view to converge to a uniform random draw of all participating nodes over time.
The main ideas behind Brahms are to use push-pull gossip-based membership with some additional defense mechanisms to prevent local views from being exclusively composed of Byzantine IDs and correct biased views due to attacks. 
We briefly describe the two components of \brahms employed by each node before detailing its defense mechanisms.

\brahms has two components.
The first is a \emph{gossip} component.
Each node spreads locally known IDs across the system and maintains a dynamic view $V$ of $l_1$ entries.
The second is a \emph{local sampling} component that enables each node to maintain a sample list $S$ of $l_2$ entries uniformly sampled from the received IDs. 
Uniformity is obtained by a min-wise independent permutation~\cite{broderMinWise2000} technique.

\begin{figure}[!htb]
  \centering
  \includegraphics[width=.75\linewidth]{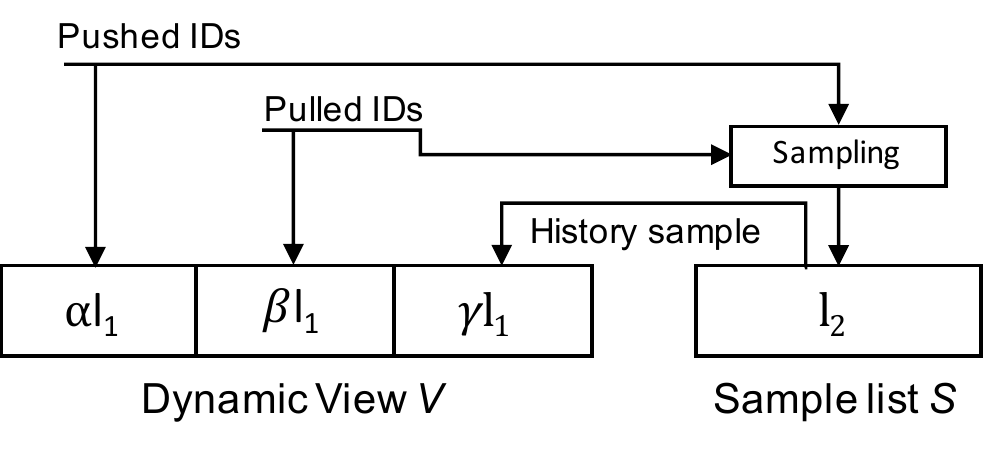}
  \caption{\brahms view computation}
  \label{fig:brahms_view}
\end{figure}

Initially, each node possesses a list containing node IDs and addresses obtained from a bootstrap node. 
Periodically, a \brahms node, through the gossip component, selects from its dynamic view $V$ both, $\alpha \times l_1$ nodes to send them push messages containing its own ID, and $\beta \times l_1$ nodes to send them pull-request messages in order to retrieve their views.
At the end of each round, the sample list $S$ is updated via the sampling component, while the dynamic view $V$ is renewed by randomly selecting: (i) $\alpha \times l_1$ IDs received from push messages, (ii) $\beta \times l_1$ IDs from pull answers, and (iii) $\gamma \times l_1$ IDs from the sample list (referred to as the history sample), with $\alpha+\beta+\gamma = 1$, as shown in Figure~\ref{fig:brahms_view}.

\begin{figure}[!htb]
  \centering
  \includegraphics[width=.85\linewidth]{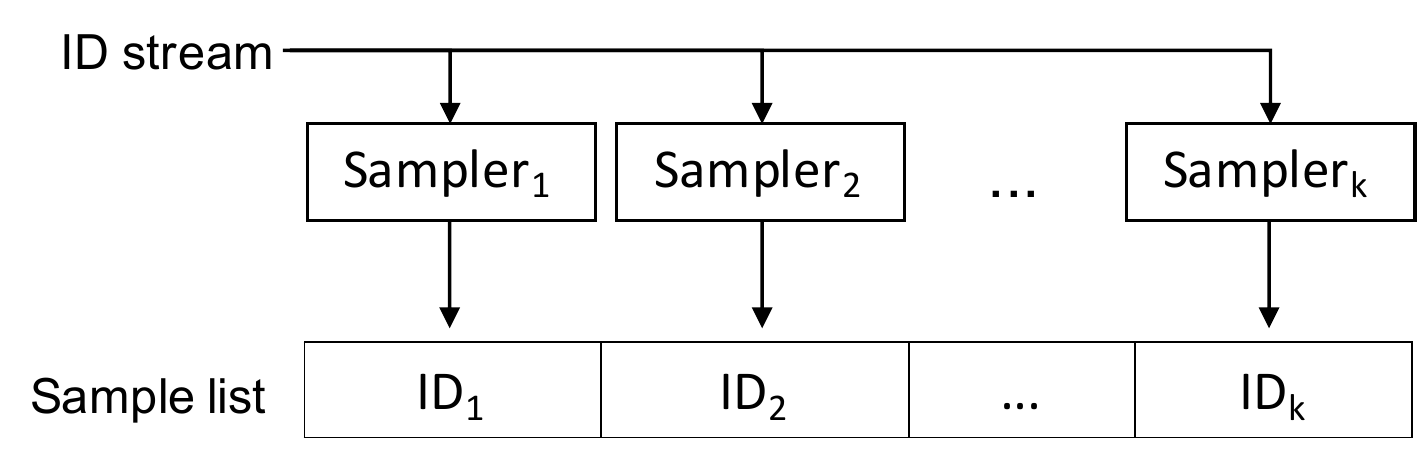}
  \caption{\brahms sampling component}
  \label{fig:brahmssampler}
\end{figure}

As depicted in Figure~\ref{fig:brahmssampler}, the sampling component consists  of $l_2$ samplers. 
It takes as input a stream of identifiers received from push or pull messages and produces a sample list of size $l_2$.
At initialization, each sampler chooses a hash function at random.
A sampler takes an identifier as input and computes its hash. 
If the calculated hash is smaller than the one corresponding to a previously stored identifier, the sampler produces the new identifier and stores it in its local memory. 
Otherwise, the sampler outputs the stored ID.


The four defense mechanisms that prevent partitioning and allow convergence to uniform sampling are the following: 
  (i)~limited pushes,
  (ii)~attack detection and blocking,
  (iii)~controlling the contribution of pulls versus pushes, and 
  (iv)~history sampling.
We detail these mechanisms in the following.
(i)~An adversary could forge identities or flood the system with push requests, leaving correct IDs propagated mainly through pulls and diminishing their representation exponentially.
\brahms assumes a mechanism that limits the message sending rate of nodes, for example, via computational challenges like Merkle's puzzles, virtual currency, \etc
(ii)~Limiting push messages prevents a simultaneous attack on all correct nodes but does not protect against flooding a targeted node. 
To do so, \brahms blocks dynamic view updates if more than the expected $\alpha \times l_1$ pushes are received.
This policy slows down progress but its expected impact in the absence of attacks is bounded, and thanks to limited pushes, some nodes make progress even under attack.
(iii)~Node views are threatened by pulls from neighbors more than by adversarial pushes.
Pushes from correct nodes are correct, but pull answers from correct nodes may contain some identifiers of Byzantine nodes.
Hence, the contribution of pushes and pulls to $V$ must be balanced. 
Brahms updates $V$ with randomly chosen $\alpha \times l_1$ pushed IDs to protect targeted nodes, and $\beta \times l_1$ pulled IDs to protect the rest.
(iv)~The attack detection and blocking technique slows down a targeted attack but cannot prevent it completely. 
A node victim of targeted pushes will pull more IDs from Byzantine nodes, send fewer pushes to correct ones, causing its systemwide representation to decrease hence receiving fewer correct pushes.
Brahms overcomes such an attack using a self-healing mechanism by incorporating, in the view, an unbiased historical sample of $\gamma \times l_1$ IDs from the sample list $S$. 
Once some correct ID becomes  the permanent sample of the node under attack (or the node’s ID becomes a permanent sample of another  correct node), the threat of isolation is eliminated. 

\begin{figure}[!htb]
  \centering
  \subfloat{
    \includegraphics[width=\linewidth]{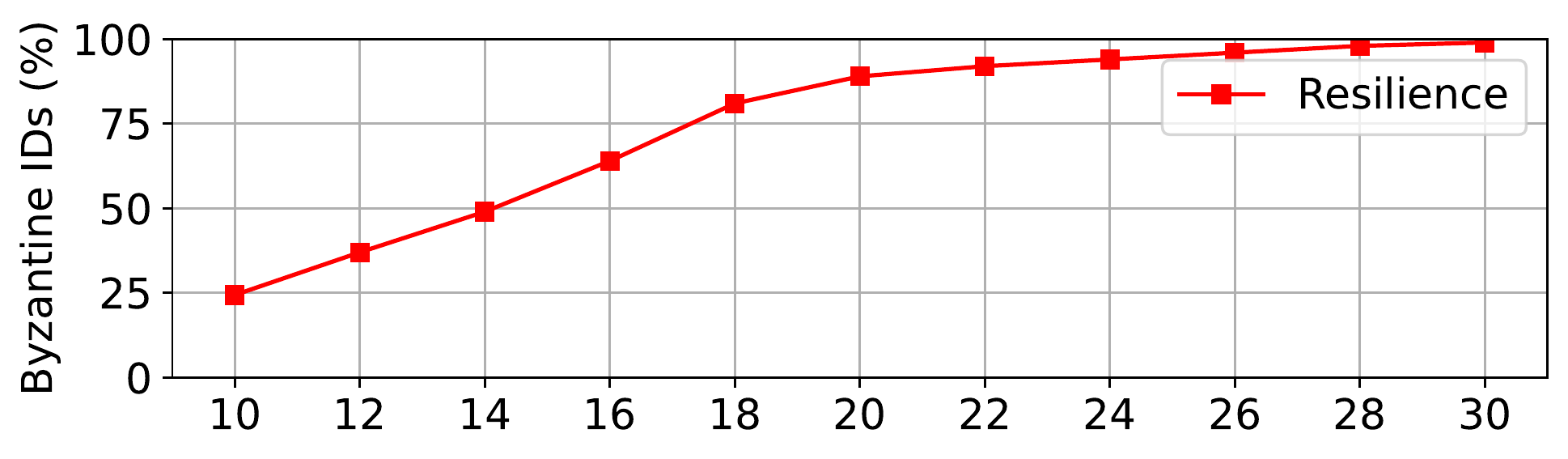}
    }
    
    \vspace{-0.5cm}

    \subfloat{
    \includegraphics[width=\linewidth]{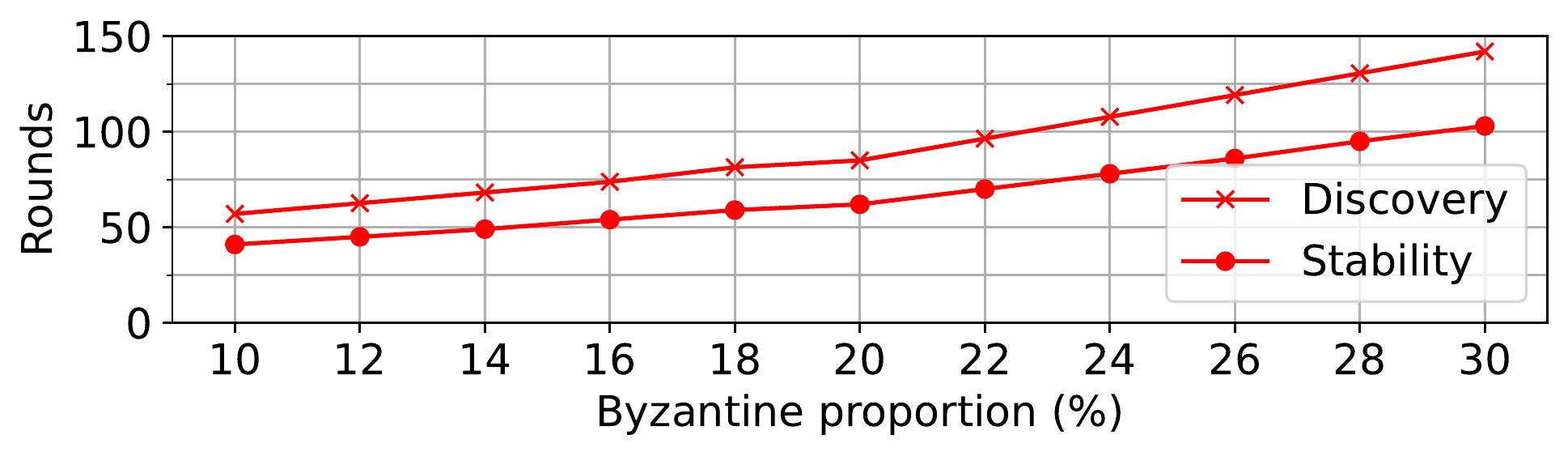}
    }
  \caption{\brahms resilience, time to discovery and to stability under Byzantine faults}
  \label{fig:brahms_baseline}
\end{figure}

Results on Figure~\ref{fig:brahms_baseline} show resilience (percentage of Byzantine IDs in the views of correct nodes), time to discovery (number of rounds required for all nodes to discover at least 75\% of non-Byzantine IDs) and time to view stability (number of rounds necessary for all non-Byzantine node views to be polluted within 10\% of the average proportion of Byzantine IDs in the views non-Byzantine nodes) of \brahms under different shares of Byzantine nodes in the system. 
The Brahms parameters are set as recommended in the original paper~\cite{bortnikov2009brahms}, namely $\alpha = \beta = 0.4$ and $\gamma = 0.2$.

\section{Models and objectives}
\label{sec:system}

We present our system model and objectives, followed by our trust and adversarial models.

\subsection{System, operating models and \protocol{}’s objectives}

\begin{figure}[!htb]
  \centering
  \includegraphics[width=\linewidth]{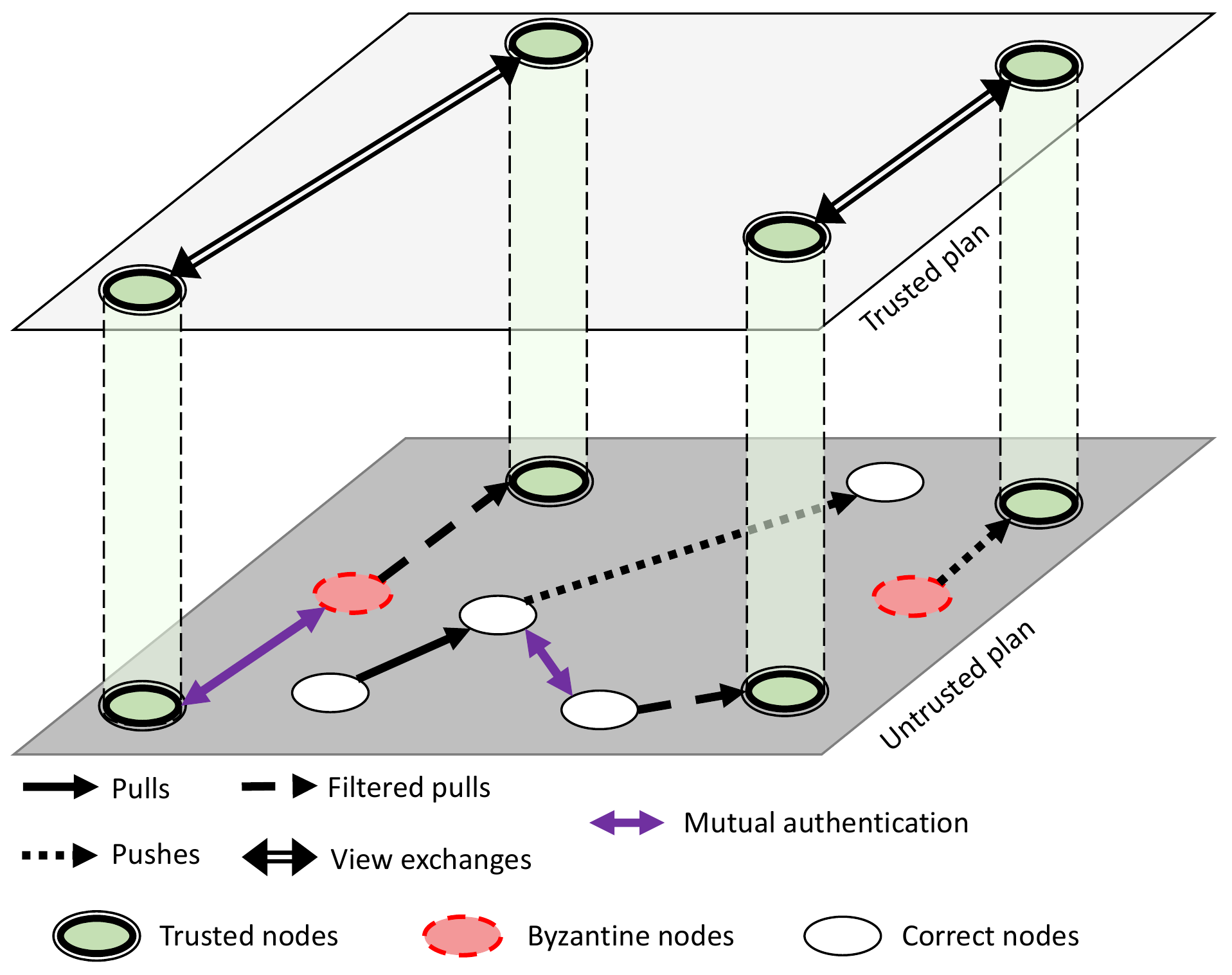}
  \caption{Overview of the \protocol protocol
  }
  \label{fig:protocol_overview} 
\end{figure}

We consider a system of $N$ nodes composed of a fraction $f$ of Byzantine nodes that can deviate from the protocol in any possible way, a fraction $t$ of trusted nodes, and a fraction $h=1-f-t$ of honest (or correct) nodes that execute \brahms. 
Each node is identified by a unique ID, chosen when the node becomes active for the first time.
Figures~\ref{fig:protocol_overview} gives an overview of \protocol and depicts the interactions between the different types of nodes. 

Honest nodes aim to get a uniform sample of the global system membership.
They execute a modified version of \brahms to exchange identifiers.
\protocol relies on both the push-pull gossip and sampling components of \brahms as detailed in Section~\ref{sec:background}, as well as its defense mechanisms.
Push messages include only the sender’s identifier, while responses to pull requests contain the full view held by the requested node.

Interactions between different types of nodes are designed to slow down the discovery of IDs possessed by untrusted nodes and accelerate the dissemination of IDs possessed by trusted nodes. 
To do so, trusted nodes filter out pull answers from untrusted ones by evicting part of the received IDs.
During trusted communications, trusted nodes exchange view parts following the instantiation of the Gossip-based Peer Sampling framework~\cite{jelasityGossipbased2007} and as detailed in the previous section.
To decide on the types of node interactions (\eg trusted to untrusted, trusted to trusted, \etc), \protocol includes a mutual authentication protocol that precedes all pull requests.

\subsection{Trust and adversarial models}

We consider an adversary controlling all Byzantine nodes with two objectives: successfully over-representing the system-wide fraction of Byzantine IDs in the partial membership knowledge of all nodes and identifying trusted nodes to evict them from the system. 
The adversary has access to the system’s global membership, including Byzantine and correct nodes, but is unaware of the location or amount of SGX-capable devices.
In the original \brahms paper~\cite{bortnikov2009brahms}, the authors prove that a balanced attack, which spreads faulty pushes evenly among correct nodes, maximizes the expected system-wide fraction of faulty IDs.
Additionally, thanks to its history sample mechanism, \brahms sustains targeted attacks where the adversary tries to partition the network by targeting a subset of nodes and sending more pushes than in a balanced attack.
Because \protocol is built on top of \brahms and inherits its properties, our adversary exclusively advertises Byzantine IDs to honest ones via answers to pull requests and evenly balanced push messages.
We identify two additional attack vectors stemming from the addition of trusted nodes into the system.
The first one consists of identifying trusted nodes to isolate them from non-Byzantine nodes, launching targeted attacks on them to pollute their views, and inserting them back into the system. 
The second one exploits the possibility of purchasing SGX-capable devices and bootstrapping their lifecycle by surrounding them with Byzantine nodes to almost fill their views with Byzantine IDs.

We rule out Sybil attacks~\cite{douceurSybil2002} by relying on the Sybil resilience of \brahms that limits the message sending rate of nodes via computational challenges like Merkle’s puzzles, virtual currency, \etc
We assume that trusted nodes can only crash fault. 
They cannot act maliciously even though Byzantine IDs can bias their view.
Our implementation uses specifically Intel SGX~\cite{costanIntel2016}. 
We trust Intel for the certification of genuine SGX-enabled CPUs, and we assume that the code running inside enclaves is properly attested before being provided with secrets.
We assume Byzantine nodes can neither break cryptographic primitives nor read data available in the trusted environment of SGX-capable devices. 
Communications between any two nodes, including trusted ones, are cyphered with symmetric encryption to protect against an eavesdropping adversary. 
We consider that an adversary cannot listen anywhere in the network and draw conclusions based on communication patterns such as message frequency or size.

\section{Interoperating trusted communications with Byzantine resilient peer sampling}
\label{sec:protocol}
We detail in this section the main components of \protocol: mutual
authentication, which allows trusted nodes to be recognized in the
mass, trusted communication, which accelerates the dissemination of
knowledge between trusted nodes, and Byzantine eviction, which slows
down view poisoning.

\subsection{Mutual authentication}
To enable trusted nodes to know whether they are communicating with other trusted nodes, we design a secure mutual authentication protocol that is executed by all nodes before issuing a pull request to a selected neighbor in the dynamic node view.
We assume that all nodes have one symmetric secret key. 
Each untrusted node generates a random secret key during the initialization phase. Conversely, trusted nodes share a common secret key that is provisioned during the remote-attestation phase.
The mutual authentication protocol between two nodes  $A$ and  $B$ operates as follows. 
First, $A$ generates a pseudo-random number $r_A$ and sends it to $B$ as a cryptographic challenge. 
In turn, $B$ generates another pseudo-random number $r_B$, it computes the hash of the concatenation of $r_A$ and $r_B$,  $H(r_A \cdot r_B)$ and encrypts it with its own secret key obtaining $[H(r_A \cdot r_B)]_{K_{B}}$. 
Then it sends $r_B$ and $[H(r_A \cdot r_B)]_{K_{B}}$ to $A$.  
Upon receiving this response, $A$ computes $H(r_A \cdot r_B)$ and deciphers $[H(r_A \cdot r_B)]_{K_{B}}$ using its own secret key $K_{A}$.
If the two values are identical (\ie $A$ and $B$ share the same secret key), $A$ can identify $B$ as trusted.
Then, $A$ sends $[H(r_B \cdot r_A)]_{K_{A}}$ to $B$. 
Like $A$, $B$ deciphers this encrypted hash using its own secret key $K_B$ and compares it with $H(r_B \cdot r_A)$. 
If the two are equal, $B$ can also identify $A$ as trusted.






\subsection{Trusted communications}\label{sec:accelerate_trusted}


In \protocol, we aim to accelerate the dissemination of identifiers among trusted nodes to spread this knowledge to untrusted nodes.
When two trusted nodes authenticate each other in a round, they can trust the remote node not to deviate from the protocol other than through failures.
For this reason, trusted nodes perform a trusted peer-sampling phase during the current round, in which they exchange and sample identifiers in a manner different from that   described for \brahms.
During the trusted communication phase, each node follows the two following measures. 
First, it swaps half of its view with half of the remote node’s view. 
Then, it transmits the received IDs to the list of pulled IDs in the \brahms protocol.
The first measure allows trusted nodes to gossip more of the IDs they receive from other trusted nodes during the round (via pull messages).
The second takes into account the IDs transmitted by other trusted nodes to update the sample list, and  renews the dynamic view during the random selection of $\beta \times l_1$ pulled IDs.

\subsection{Byzantine eviction}

Since pull answers from Byzantine nodes contain exclusively Byzantine IDs, one could think that if trusted nodes did not send any pull requests, they could prevent the poisoning of their views.
However, this simple approach would open the door for an attacker to identify trusted nodes and eclipse them from the system.
Indeed, they would behave differently from untrusted nodes, and monitoring their messages would make them easily identifiable.
For this reason, trusted nodes send pull requests in the same way as untrusted nodes and remain hidden from the sight of an eavesdropping adversary. 

Nonetheless, they can integrate an additional defense mechanism to protect their views from view-poisoning attacks without revealing themselves to the adversary.
At the end of each round, they can ignore part of the pulled IDs from untrusted nodes by not passing them to the \brahms sampling component and by ignoring them during the renewal of the pulled $\beta \times l_1$ entries of the node's dynamic view $V$.

We refer to the proportion of ignored pulled IDs as the \textit{eviction rate}.
It can be a fixed value (between 0 and 100\%) for the entire system.
A 100\% eviction rate allows the peer sampling service to construct views as if the  trusted nodes were not issuing any pull requests.
The eviction rate can also be adaptive and local for each trusted node.
That is, its value is set according to the proportion of trusted nodes with which the trusted node has exchanged identifiers during the current round.
The intuition behind adapting the eviction rate value is that the greater the share of trusted nodes contacted during the current round, the greater the number of IDs received from trusted nodes, and the fewer the attempts to poison their views from pull responses from Byzantine nodes. 
If a trusted node exchange messages with many other trusted nodes in a round, then it will evict a smaller share of IDs received from untrusted nodes compared to when it contacts a  few, if any, trusted nodes. 
We establish the following adaptive rule to determine the value of the eviction rate.
First, we limit its value between 20\%, when the proportion of trusted communications is above 80\%, and 80\%, when it is below 20\%.
A linear function governs the value of the eviction rate within these two limits.

Although evicting IDs from pull responses leads to fewer poisoned views of trusted nodes compared to non-Byzantine untrusted ones, it also helps an attacker identify trusted nodes. 
Indeed, Byzantine nodes can compare the composition of pull responses from the nodes they contact and isolate the responses that contain the fewest Byzantine IDs. 
We assess this identification risk in section~\ref{sec:sec}.

\section{Experimental Evaluation}
\label{sec:exp}

Our evaluation aims to answer the following research questions: (1) By how much does \protocol increase resilience to Byzantine faults compared to \brahms? (2) How much does \protocol impact the performance of \brahms in terms of convergence time to system discovery and view stability?

The implementation of \protocol used for our experiments consists of 1,200 lines of shared C++ code, between trusted and untrusted nodes.
The code that runs inside trusted nodes uses the Intel SGX SDK\footnote{Intel SGX SDK. https://software.intel.com/en-us/sgx/sdk.} and amounts to about 800 additional lines of C++ code.
Cryptographic operations use Intel’s OpenSSL SGX port\footnote{Intel SGX SSL. https://github.com/intel/intel-sgx-ssl.}, using RSA for asymmetric encryption and the AES-CTR mode for symmetric encryption. 
The code that runs inside untrusted nodes has about 300 additional lines of C++ code.
Since we do not have access to a large infrastructure supporting SGX-capable devices, we first perform a micro-benchmark to evaluate the overhead of running the code of \protocol's trusted node in a real SGX environment compared to an emulated SGX environment.
It enables us next to calibrate our emulated SGX environment to conduct a representative large-scale experiment involving 10,000 nodes using the Grid 5000 testbed.

\subsection{Overhead of using SGX nodes}\label{sec:microbenchmark}


Our first set of experiments are performed on a cluster of 40 physical machines.
Each machine is an Intel Next Unit of Computing (NUC) Kit with a 2-core 3.50 GHz Intel i7 processor and 32 GB of RAM\footnote{One of the models recommended by Intel for experimenting with SGX.}.
We consider a testbed of 200 nodes, with five nodes per machine.
We instrumented the \protocol's code that runs inside the trusted nodes to  measure the CPU-cycle consumption of each of the following functions: pull requests, push messages, trusted communication, sample list computation, dynamic view computation.
We also implemented a modified version of the trusted code that emulates SGX and runs on non-capable SGX devices. 
We then compare the execution time of these functions on both capable and non-capable SGX devices. 

We run two sets of experiments.
In the first set, we deploy 100 untrusted nodes and 100 trusted nodes whose code actually uses SGX capabilities.
In the second set, the 100 trusted nodes use the emulated version of the code instead.
To start the experiment, each node initiates \protocol with a view composed of a uniform random sample of the global membership.
A single experiment lasts for 200 rounds of 2.5 seconds each.
We empirically set the number of rounds to 200 to reach the convergence of the protocol.
We repeat each experiment 100 times and collect measurements at each iteration. We then compare the running time of operations on the emulated SGX nodes and on the real SGX ones, and compute the associated overhead. 
Table~\ref{table:overhead} shows these running times together with the  mean and standard deviation of the CPU-cycle overhead for the functions under consideration. We use this overhead to calibrate our emulated implementation in the rest of our experiments. 




\begin{table}[!htb]
  \centering
  \caption{SGX performance overhead (in CPU cycles)}
  \begin{tabular}{l r r c c}
      \toprule
      Peer sampling function  & \makecell{Standard} & \makecell{SGX} & \makecell{Mean\\overhead} & \makecell{Standard\\deviation}  \\
      \midrule
      Pull request           & 15,623 & 18,593 & 2,970 & 3 \%  \\
      Push message           &  7,521 &  9,182 & 1,661 & 3 \%  \\
      Trusted communications &  9,845 & 11,516 & 1,671 & 3 \%  \\
      Sample list comput.    & 13,024 & 15,364 & 2,340 & 4 \% \\
      Dynamic view comput.   & 12,457 & 15,076 & 2,619 & 2 \%  \\
      \bottomrule
  \end{tabular}
  \label{table:overhead}
\end{table}


\begin{figure*}[!htb]
  \centering     
  \subfloat[Byzantine resilience gain]{
    \label{fig:res0}\includegraphics[width=0.3\linewidth]{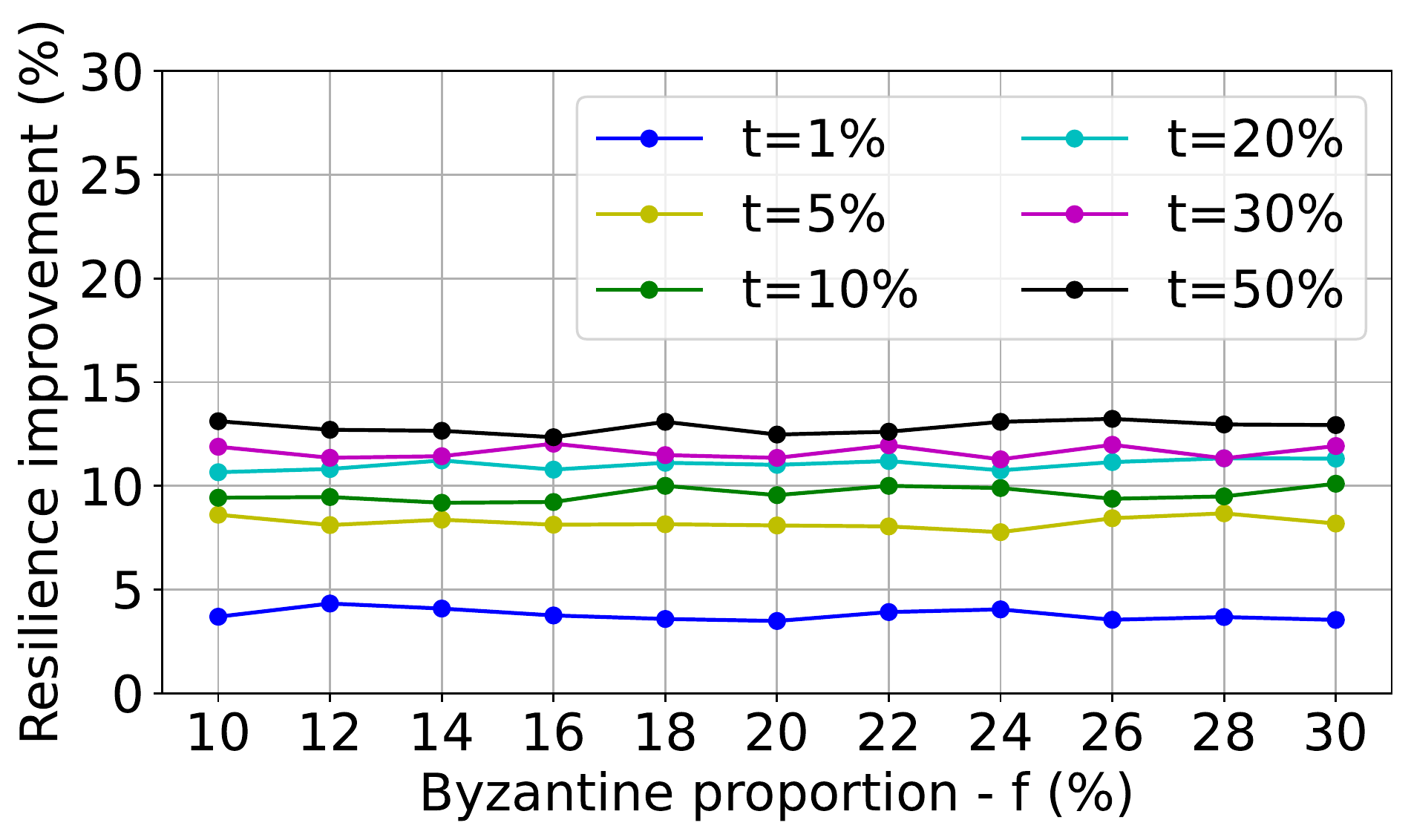}
    }
  \subfloat[Round overhead for system discovery (\%)]{
    \label{fig:couv0}\includegraphics[width=0.3\linewidth]{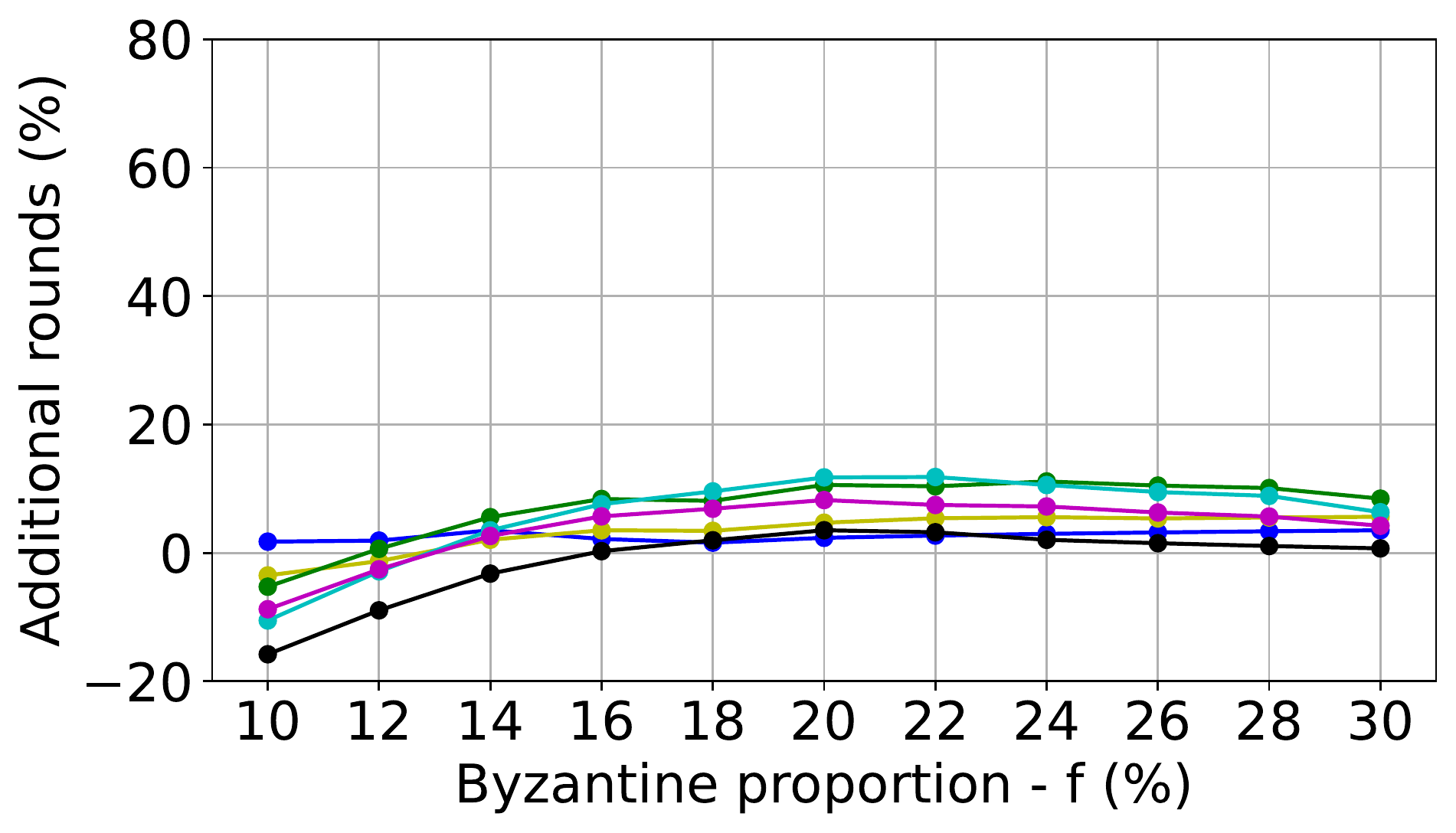}
    }
  \subfloat[Round overhead to reach view stability (\%)]{
    \label{fig:stab0}\includegraphics[width=0.3\linewidth]{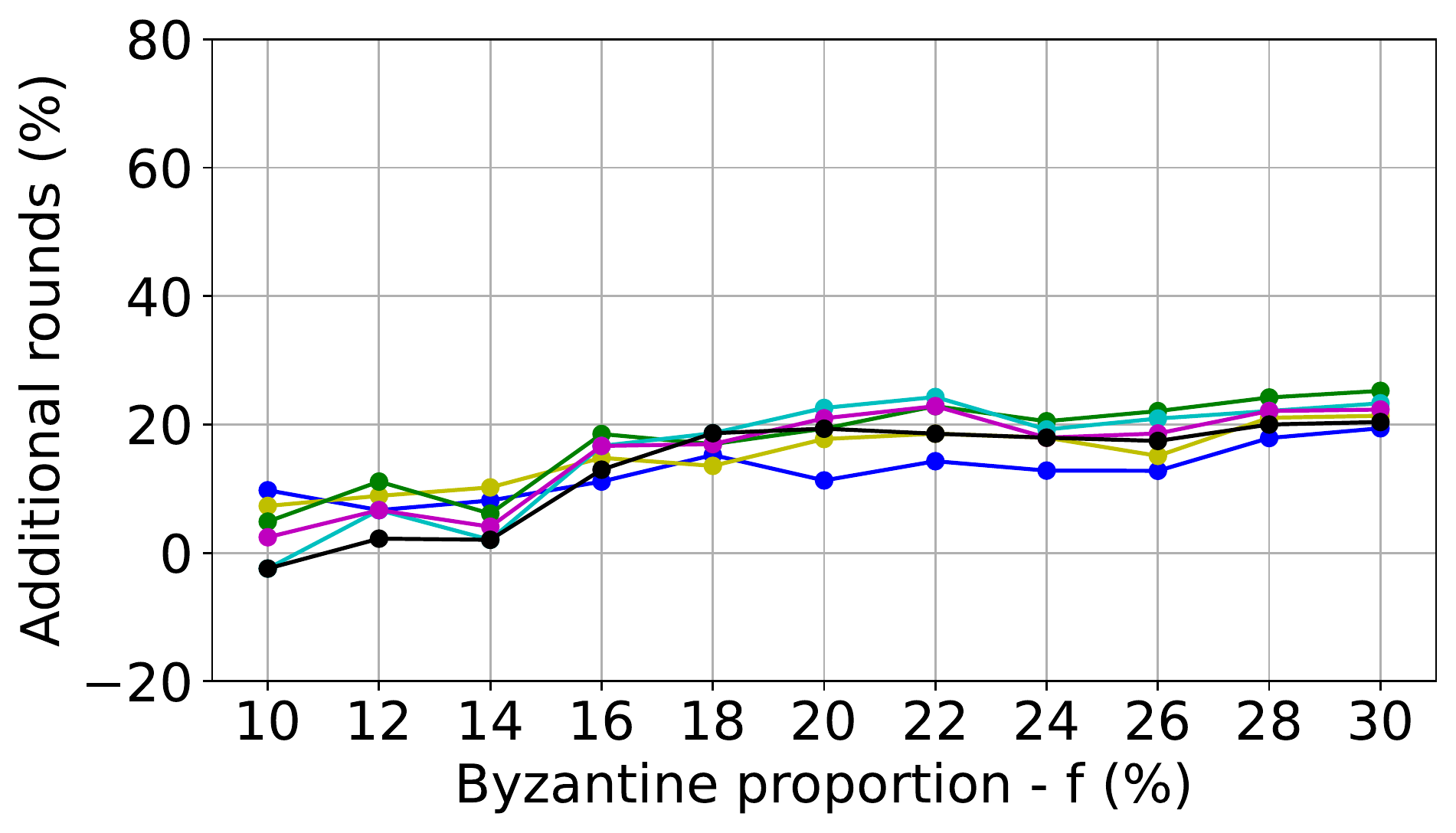}
  }
  \caption{Resilience improvement and performance overhead under a 0\% eviction rate\vspace{-0.1cm}}
  \label{fig:0}
  \centering     
  \subfloat[Byzantine resilience gain]{
    \label{fig:res40}\includegraphics[width=0.3\linewidth]{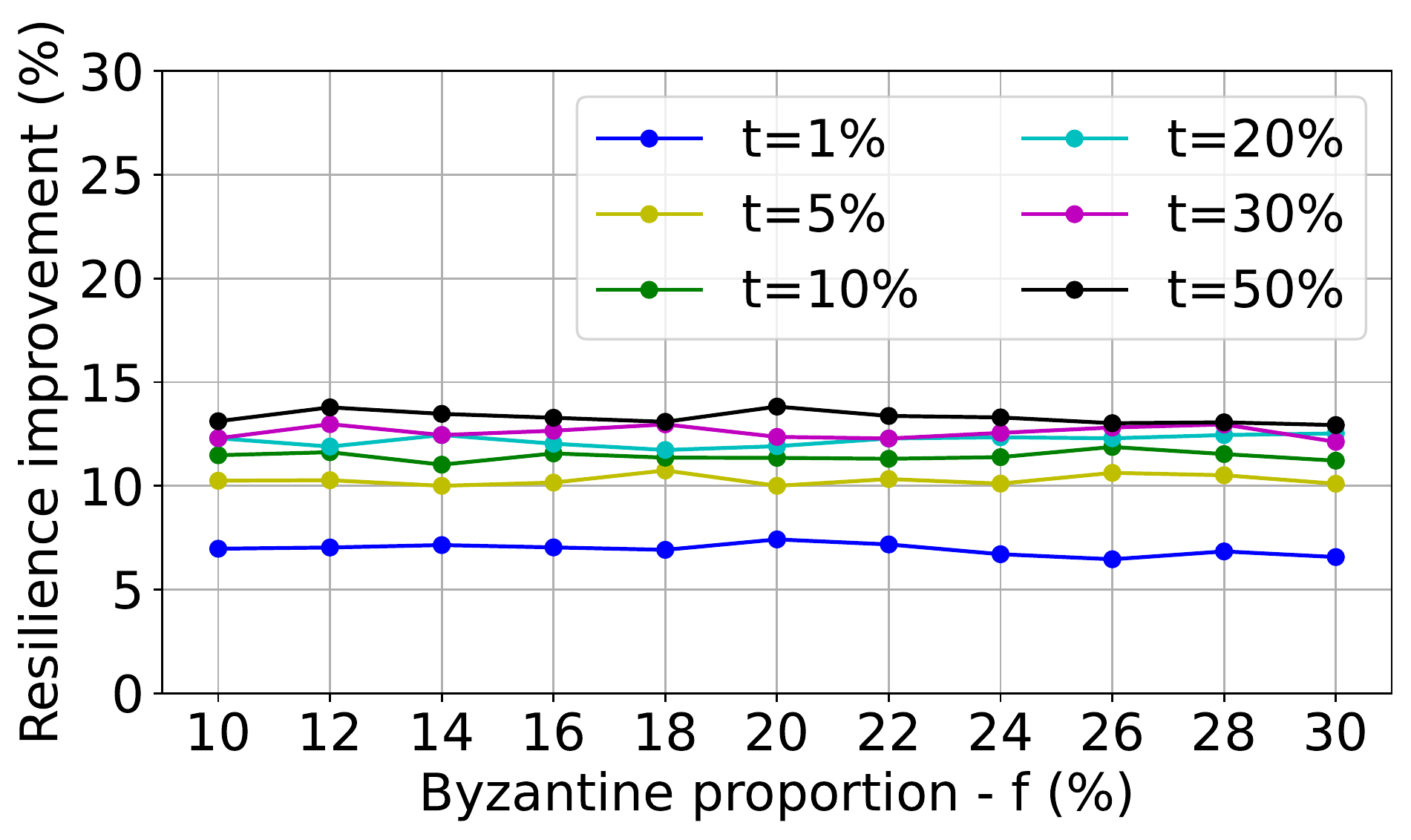}
    }
  \subfloat[Round overhead for system discovery (\%)]{
    \label{fig:couv40}\includegraphics[width=0.3\linewidth]{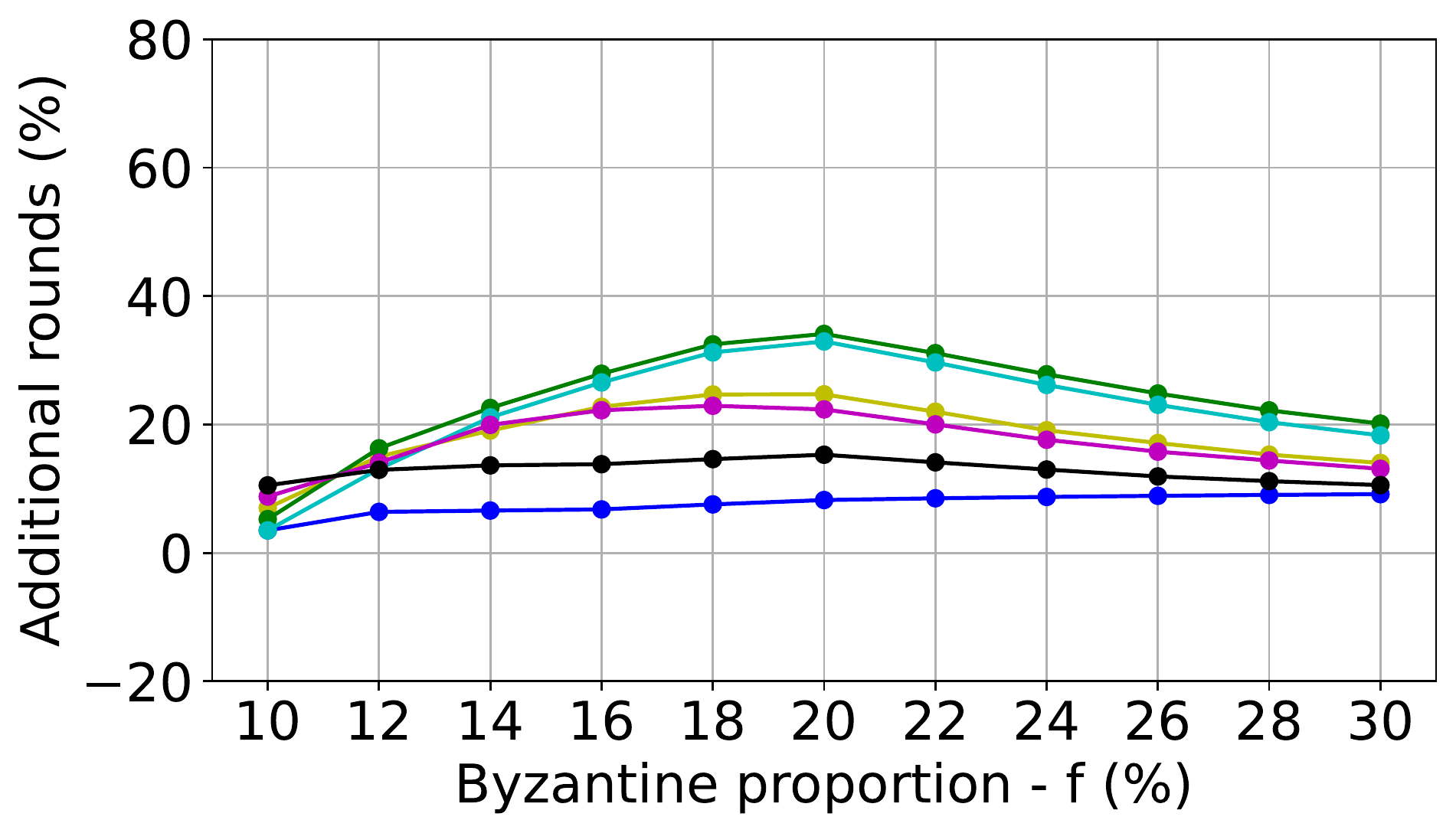}
    }
  \subfloat[Round overhead to reach view stability (\%)]{
    \label{fig:stab40}\includegraphics[width=0.3\linewidth]{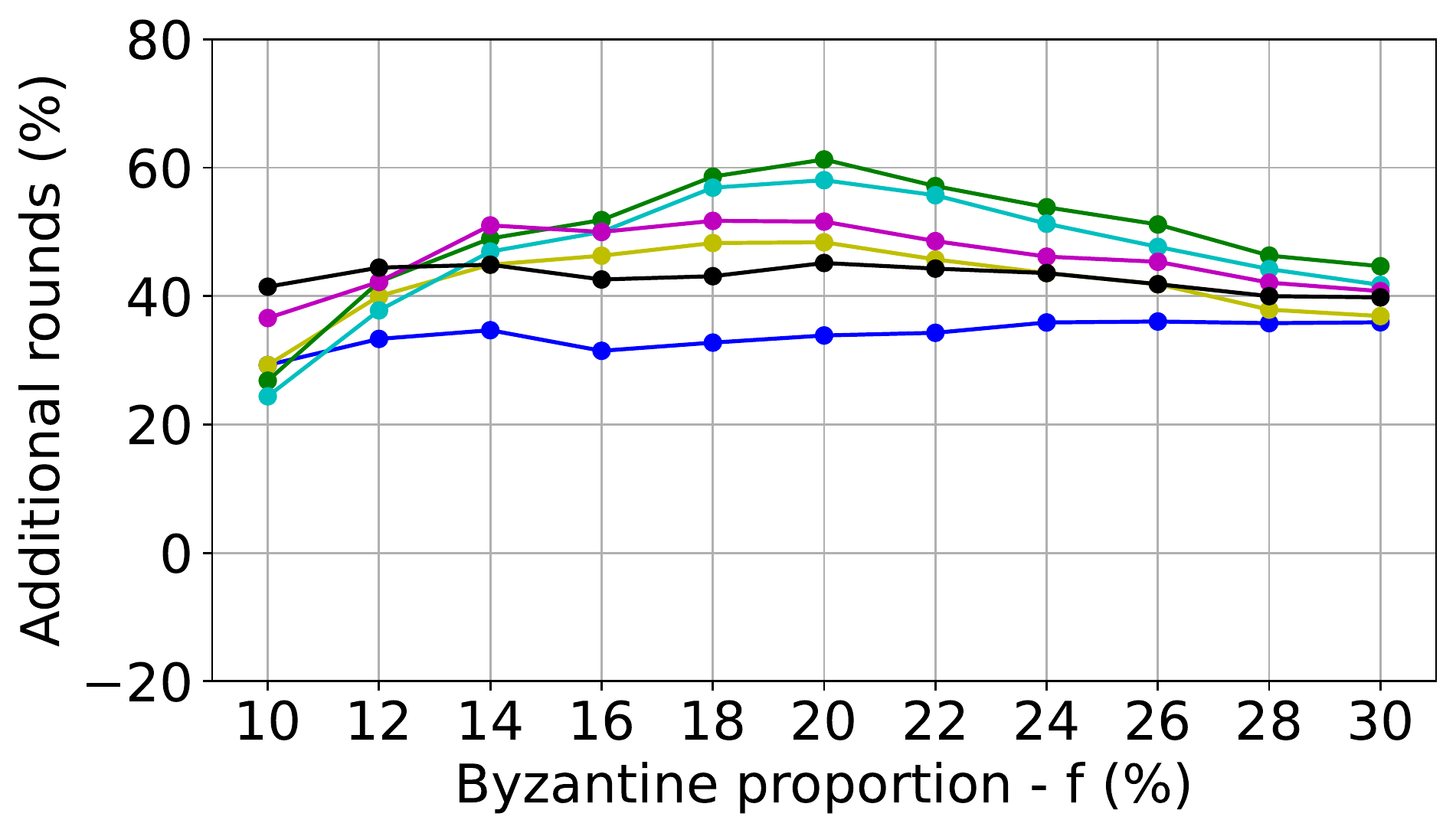}
  }
  \caption{Resilience improvement and performance overhead under a 40\% eviction rate\vspace{-0.1cm}}
  \label{fig:40}
  \centering     
  \subfloat[Byzantine resilience gain]{
    \label{fig:res60}\includegraphics[width=0.3\linewidth]{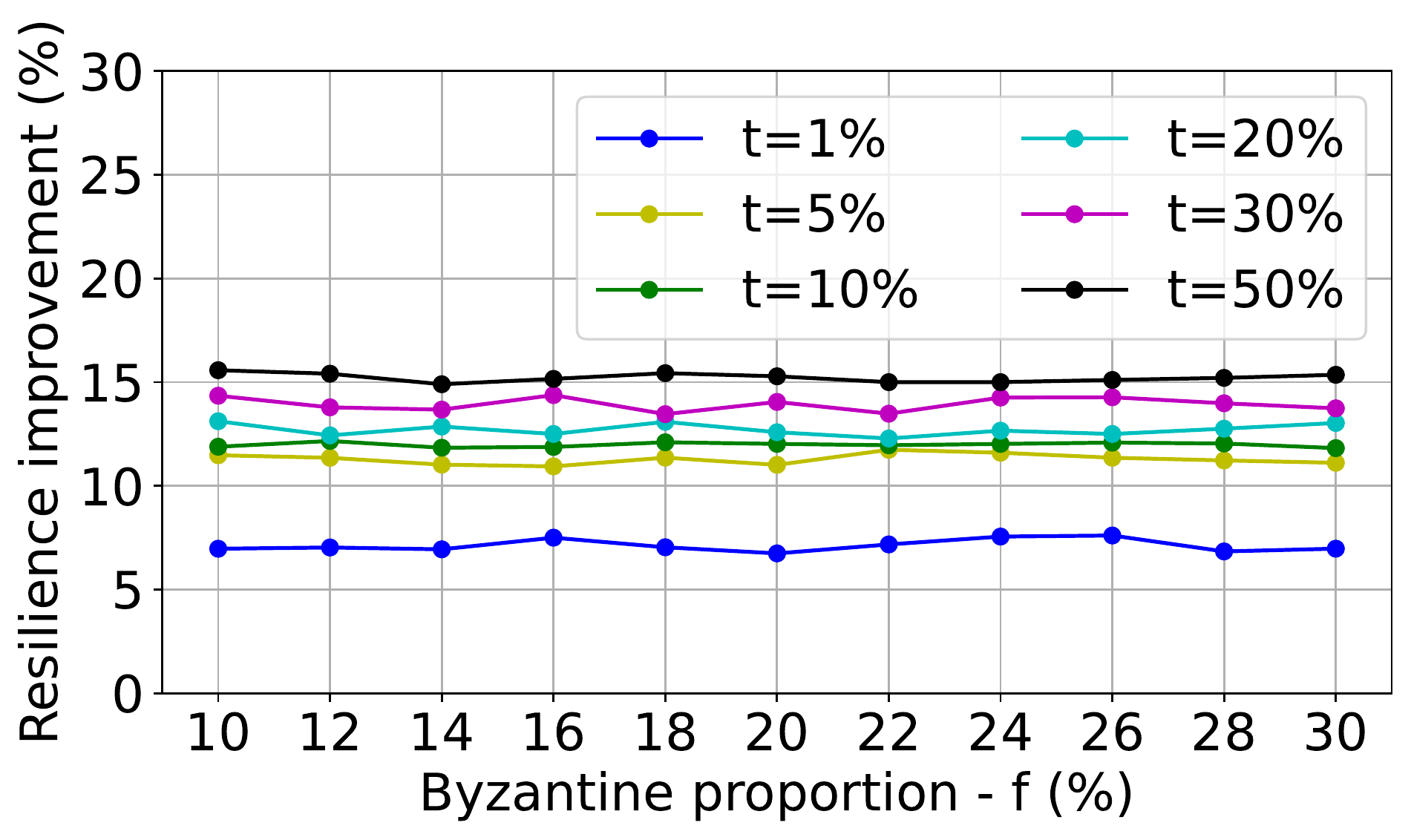}
    }
  \subfloat[Round overhead for system discovery (\%)]{
    \label{fig:couv60}\includegraphics[width=0.3\linewidth]{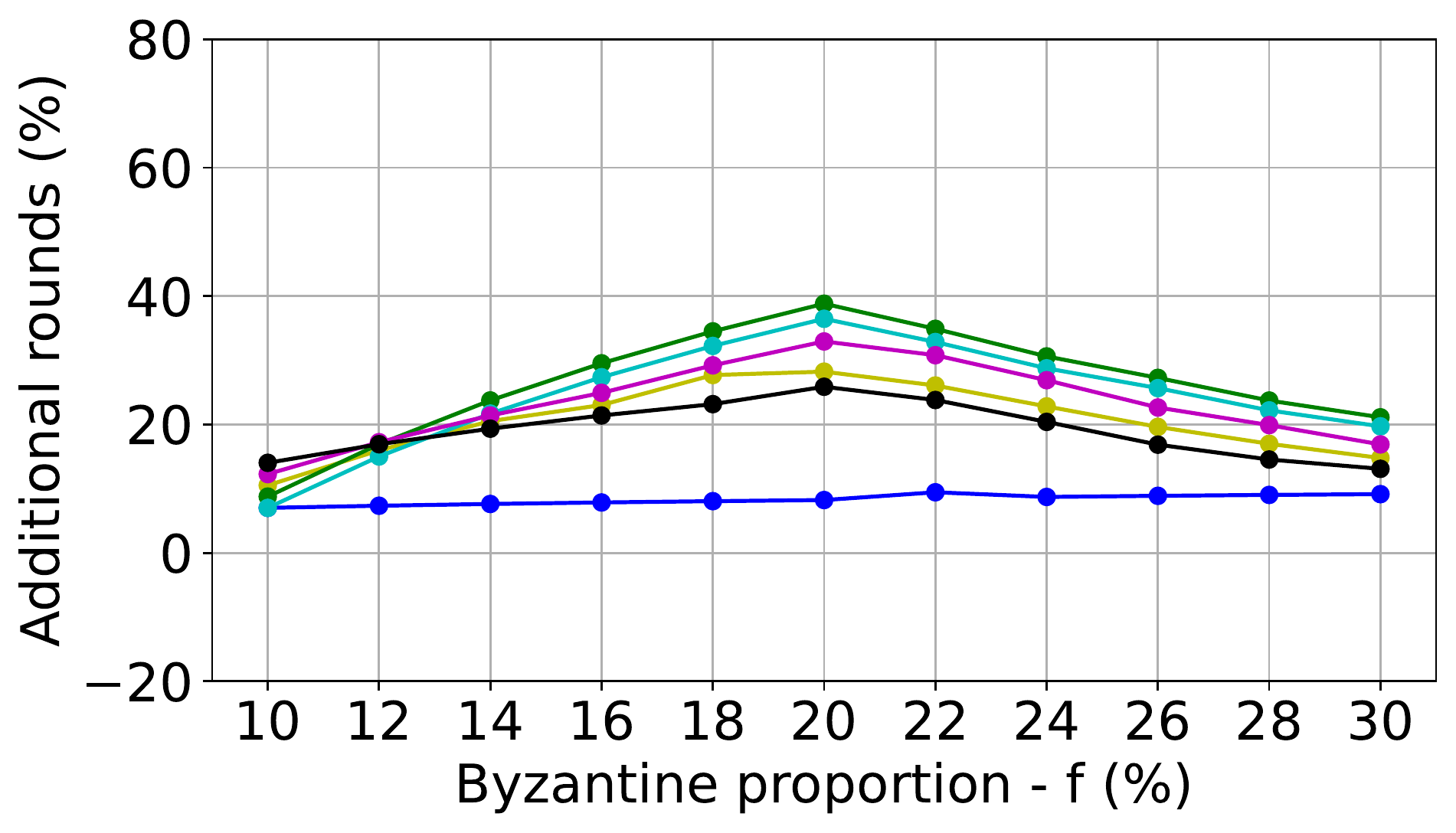}
    }
  \subfloat[Round overhead to reach view stability (\%)]{
    \label{fig:stab60}\includegraphics[width=0.3\linewidth]{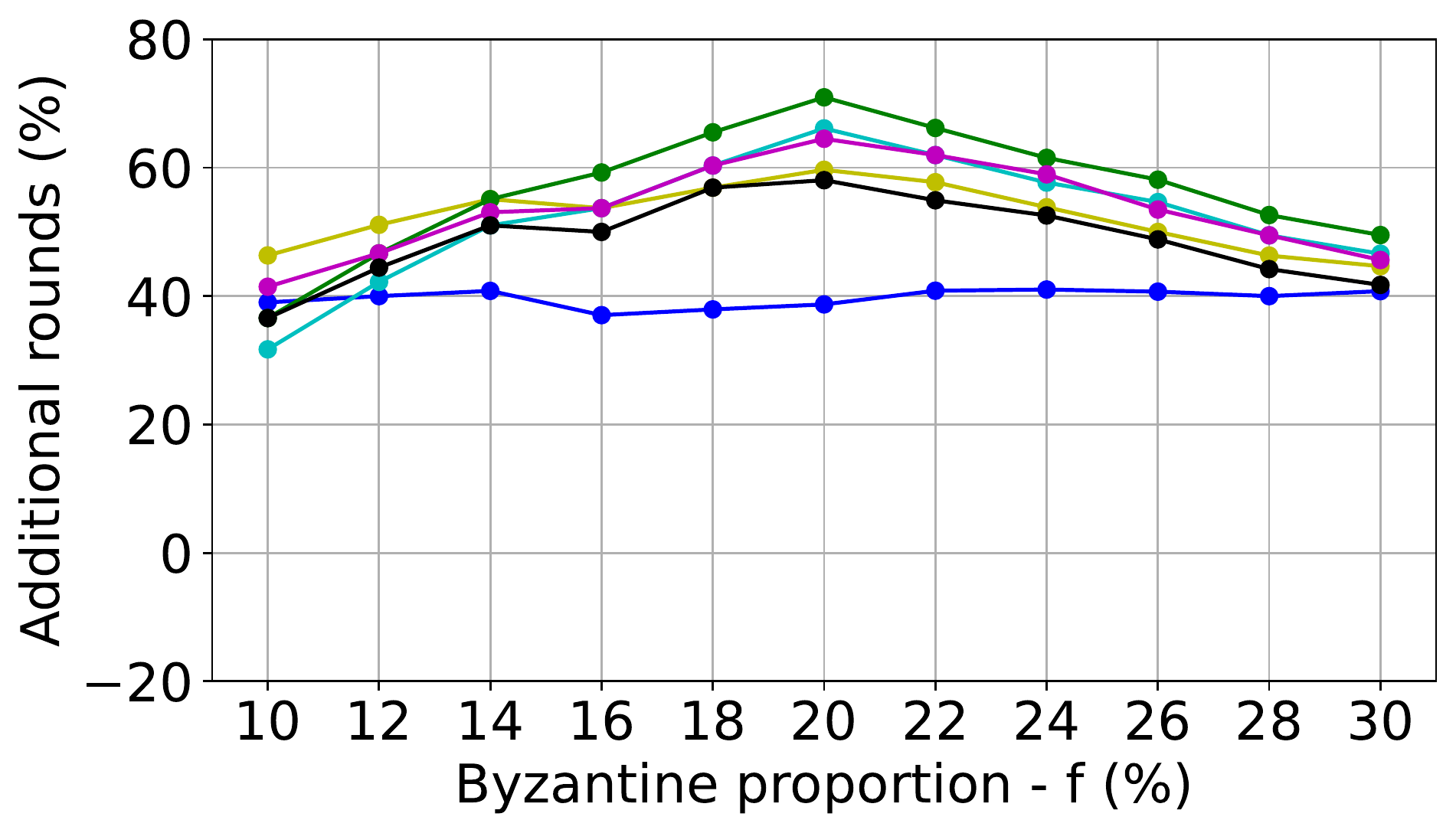}
  }
  \caption{Resilience improvement and performance overhead under a 60\% eviction rate\vspace{-0.1cm}}
  \label{fig:60}
\end{figure*}

\begin{figure*}
  \centering     
  \subfloat[Byzantine resilience gain]{
    \label{fig:res100}\includegraphics[width=0.3\linewidth]{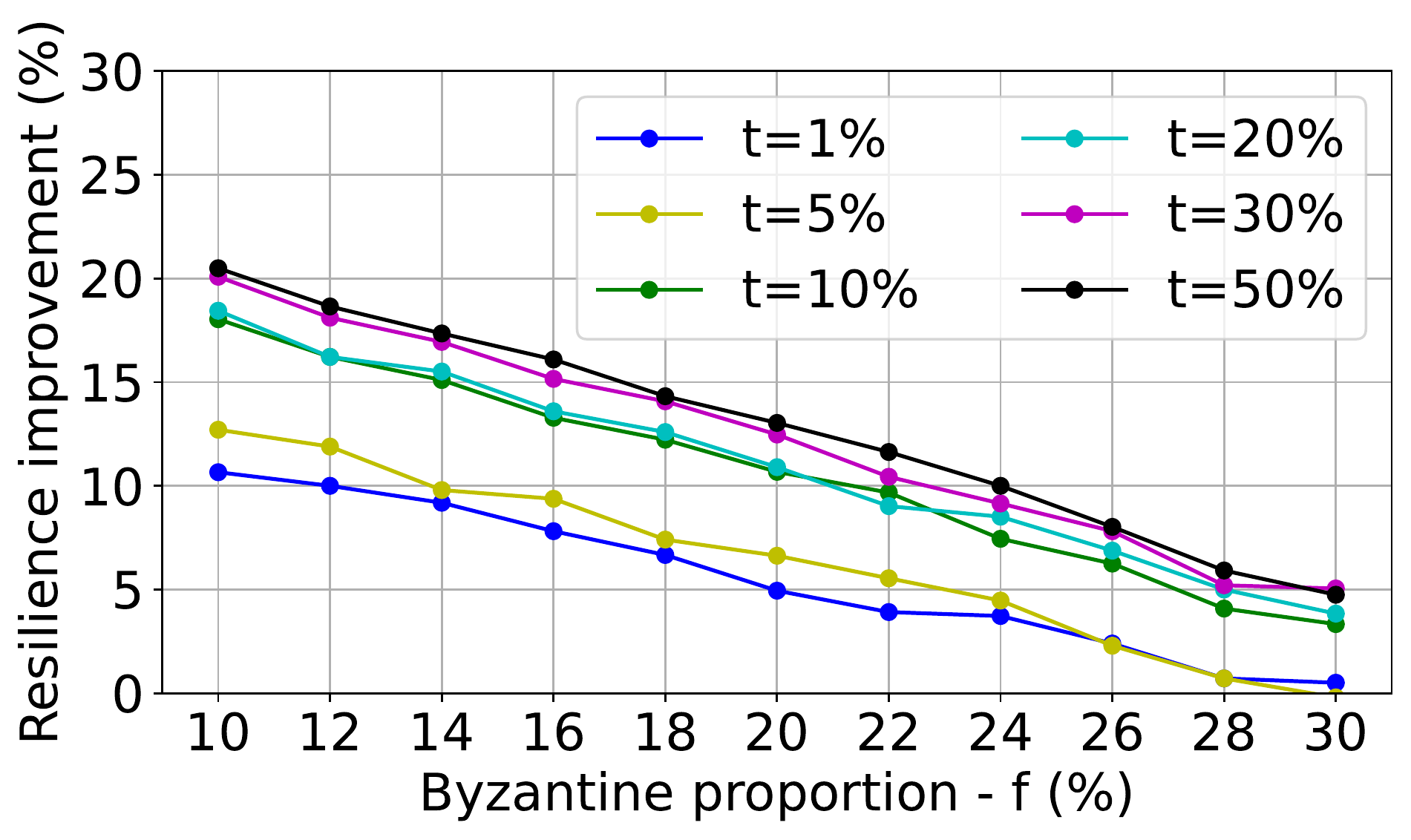}
  }
  \subfloat[Round overhead for system discovery (\%)]{
    \label{fig:couv100}\includegraphics[width=0.3\linewidth]{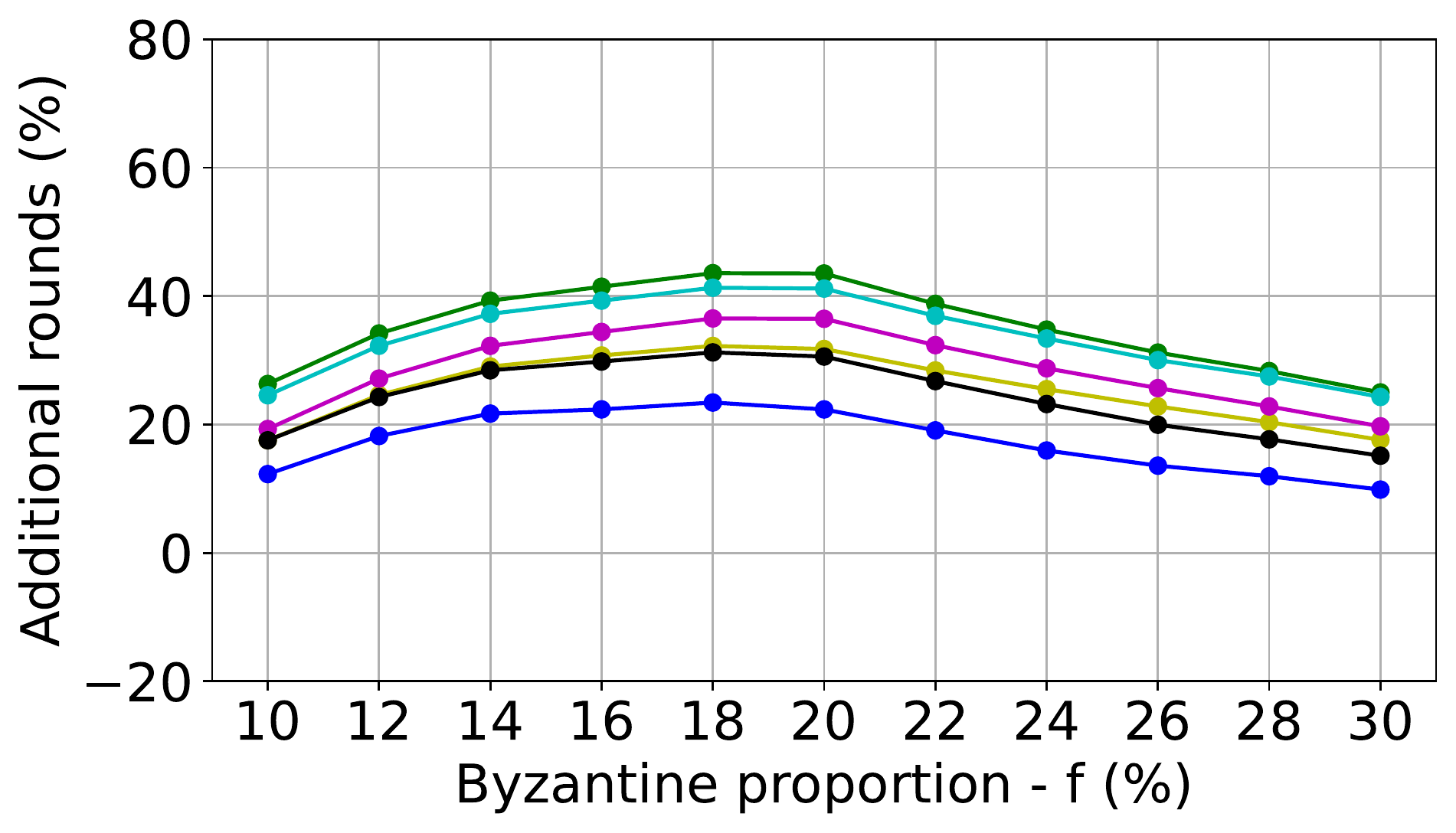}
    }
  \subfloat[Round overhead to reach view stability (\%)]{
    \label{fig:stab100}\includegraphics[width=0.3\linewidth]{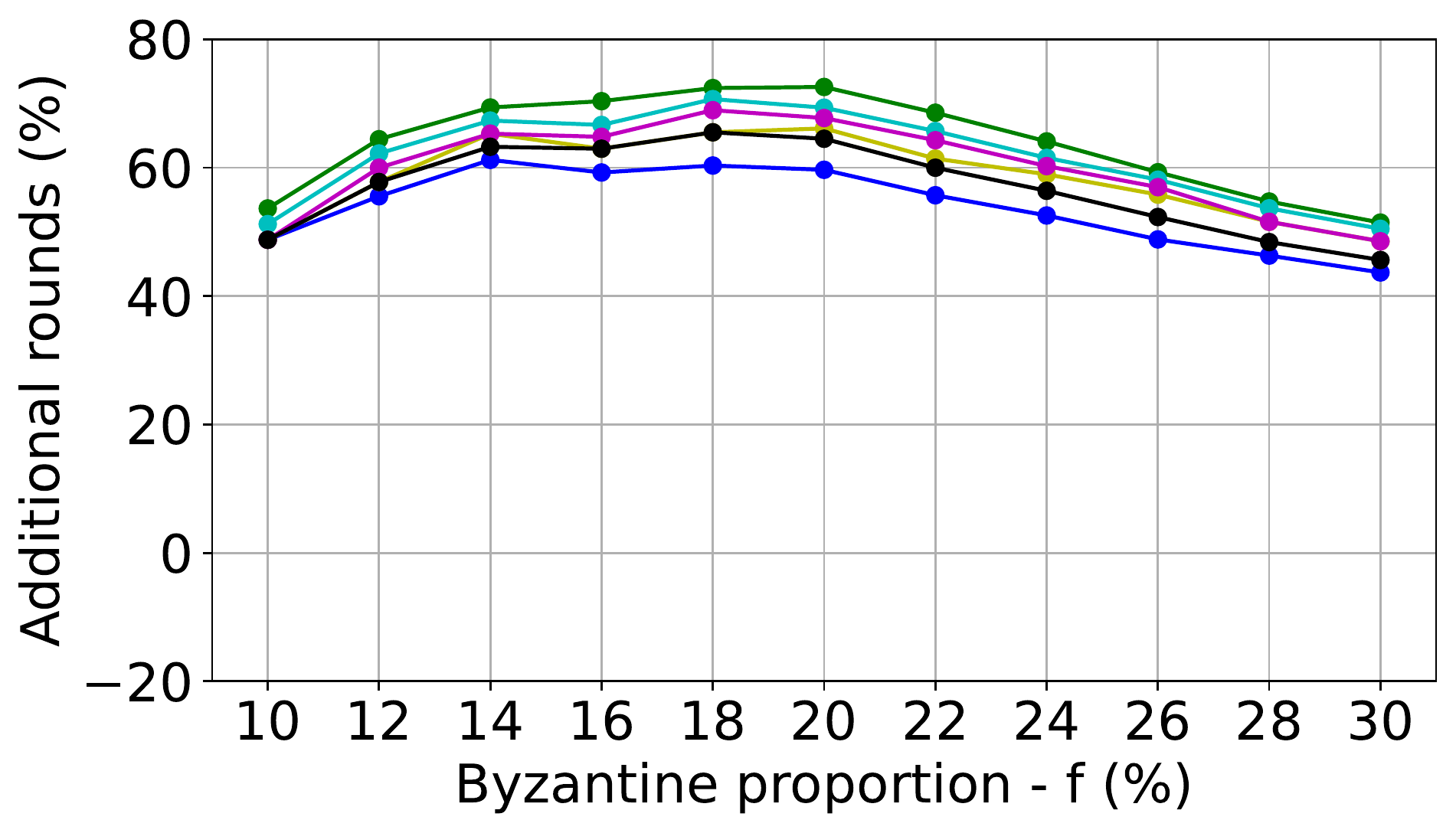}
  }
  \caption{Resilience improvement and performance overhead under a 100\% eviction rate\vspace{-0.1cm}}
  \label{fig:100}

  \centering     
  \subfloat[Byzantine resilience gain]{
    \label{fig:resadapt}\includegraphics[width=0.3\linewidth]{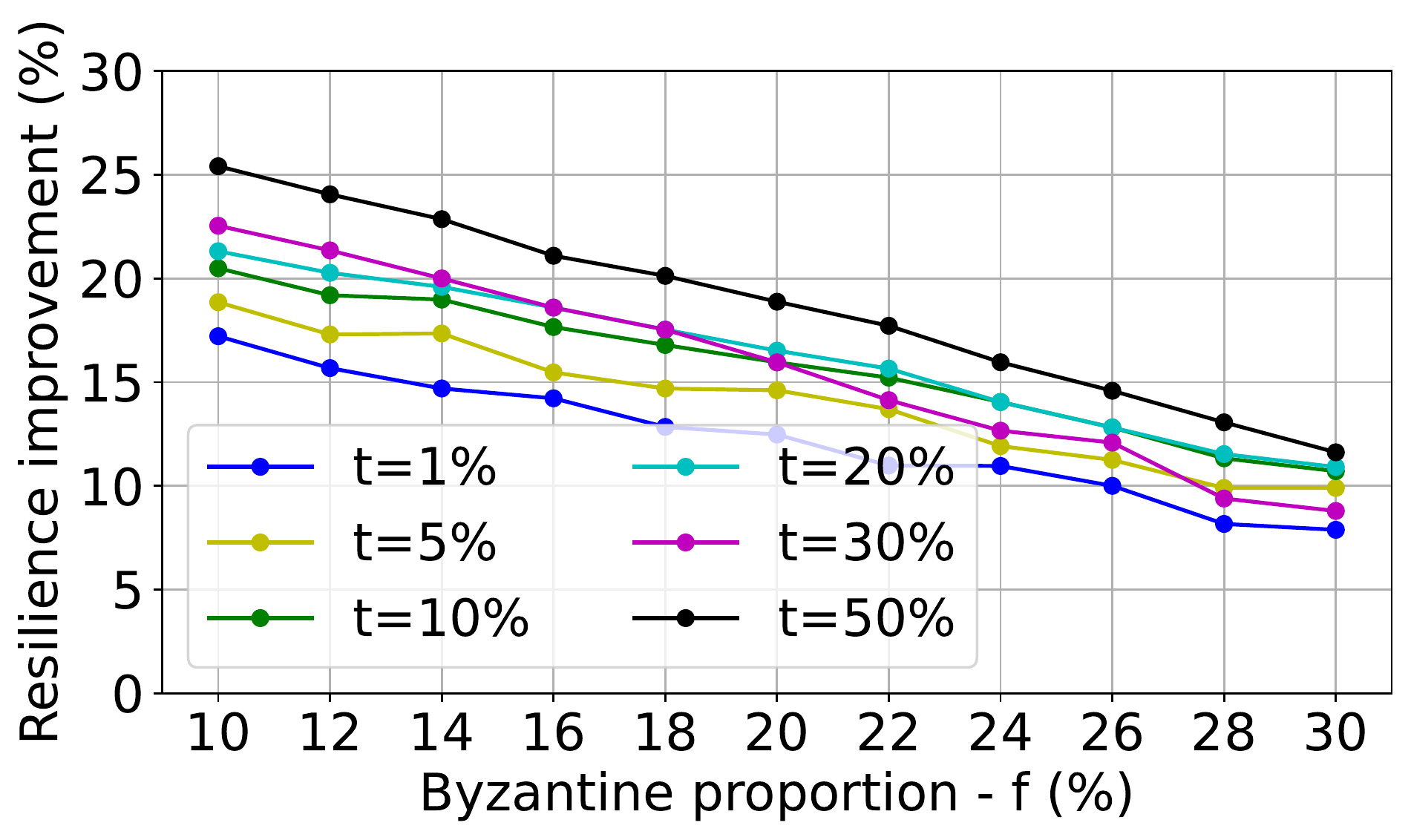}
    }
    \subfloat[Round overhead for system discovery (\%)]{
      \label{fig:couvadapt}\includegraphics[width=0.3\linewidth]{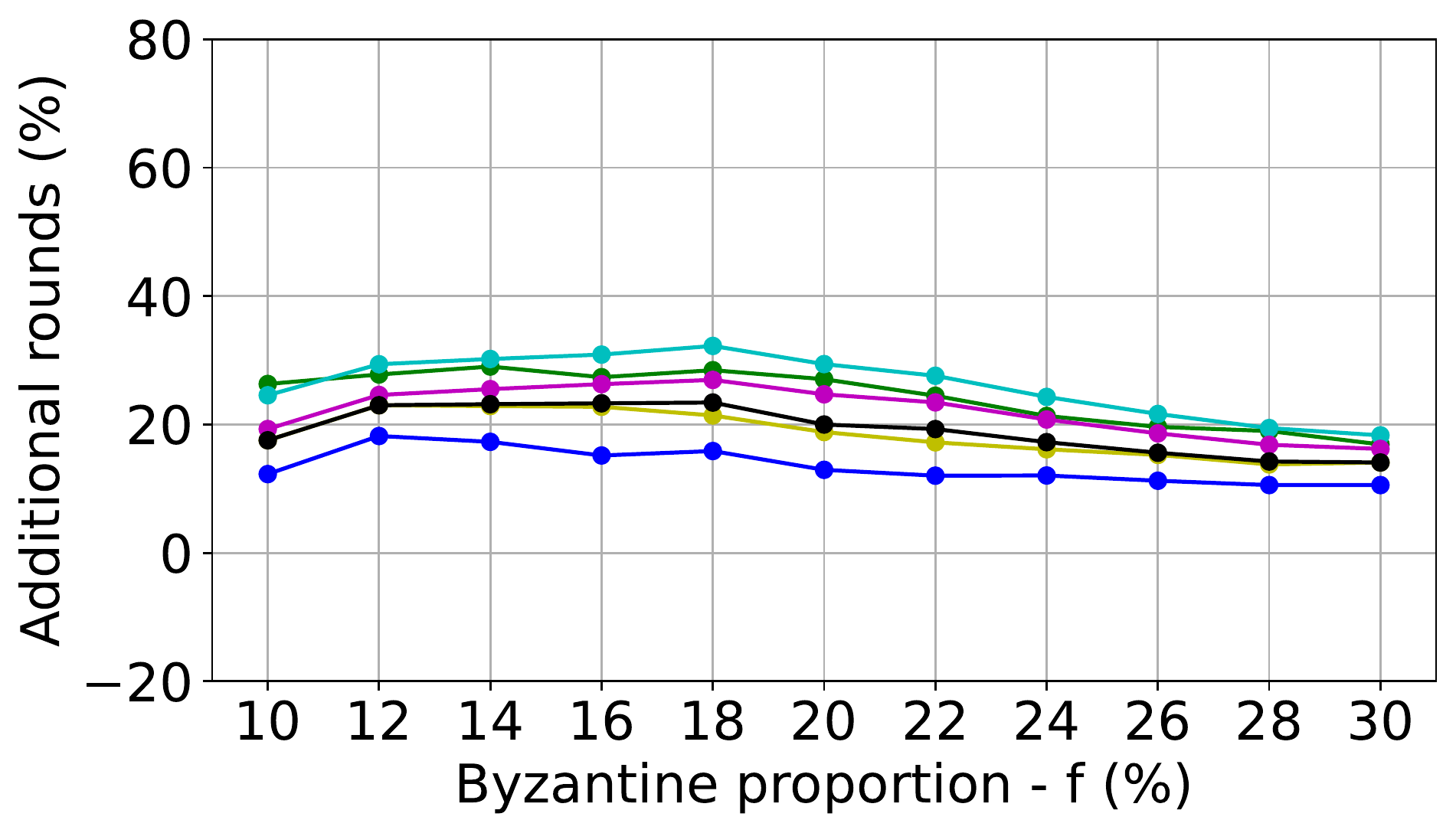}
      }
    \subfloat[Round overhead to reach view stability (\%)]{
      \label{fig:stabadapt}\includegraphics[width=0.3\linewidth]{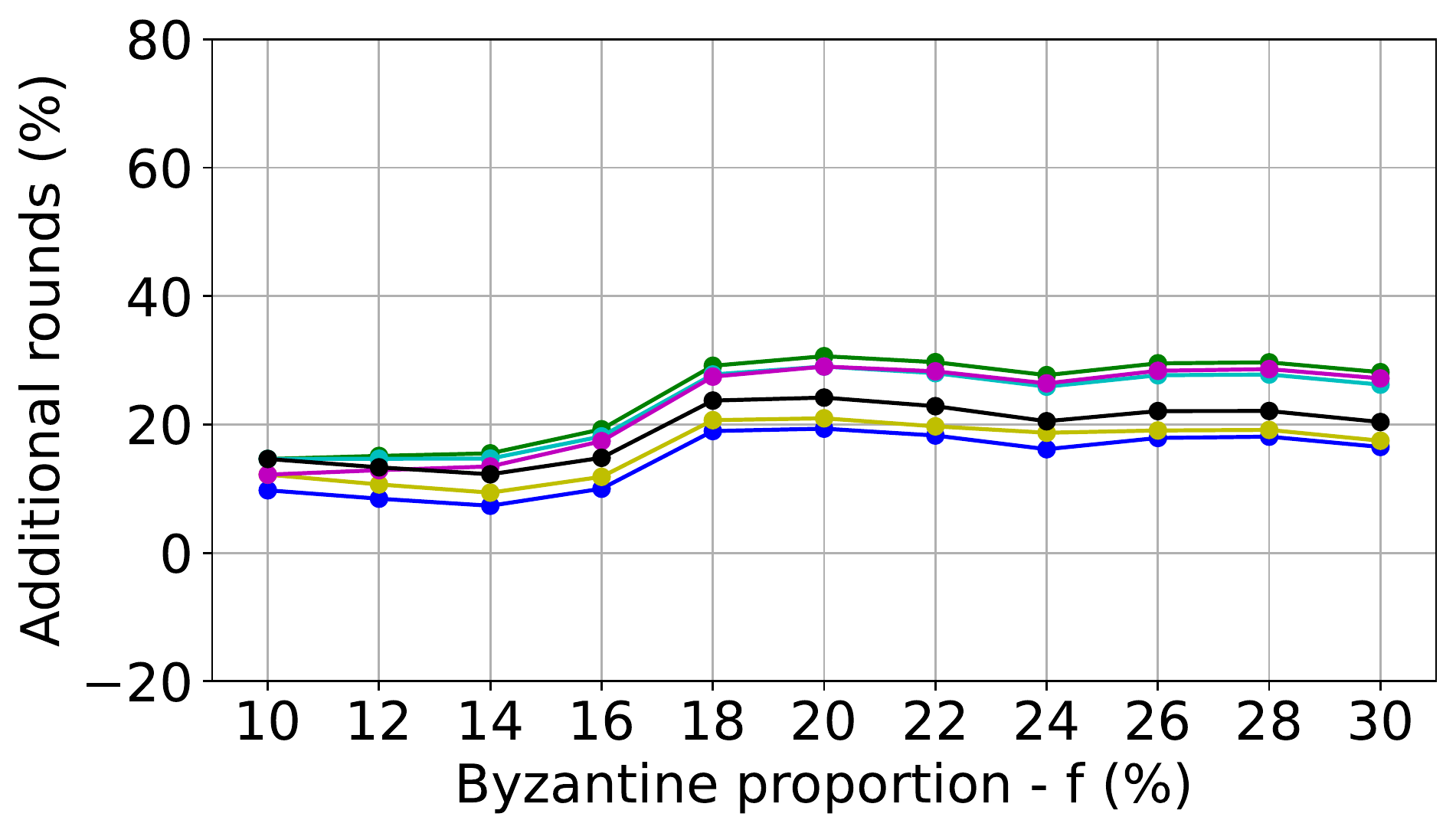}
    }
  \caption{Resilience improvement and performance overhead under the adaptive eviction rate policy\vspace{-0.1cm}}
  \label{fig:adapt}
\end{figure*}

\subsection{Large-scale experimental setup}

We emulate the behavior of SGX-capable nodes in a larger testbed
hosted on Grid 5000.  We increase the latency of each function that
needs to be executed in SGX by adding a random delay that depends on
the mean CPU-cycle overhead and follows its standard deviation, based
on the data in
Table~\ref{table:overhead}. 
Our emulated testbed consists of 10,000 nodes, running on 20 machines.
Each machine has 2 CPUs with 16 cores each and 192~GB of memory.

We evaluate the resilience of \protocol and compare it to the nominal value of \brahms (see Section~\ref{sec:background}).
Resilience improvement is defined as the percentage drop in the number of Byzantine identifiers in the views of correct nodes.
We also evaluate the performance of \protocol compared to \brahms on two other indicators: system-discovery time and view-stability time.
As in the case of \brahms (Section~\ref{sec:background}), system-discovery time is defined as the number of rounds required for all nodes to discover at least 75\% of non-Byzantine IDs.
View-stability time is defined as the number of rounds necessary for all non-Byzantine node views to be polluted within 10\% of the average proportion of Byzantine IDs in the views of non-Byzantine nodes.
We measure the performance overhead in terms of a percentage of additional rounds.

Each node runs an experiment for 200 rounds of 2.5-seconds each, during which it records the composition of its view in terms of Byzantine and honest IDs.
We repeat each experiment 10 times to consolidate the results.
An experimental setup consists of selected proportions of Byzantine nodes, $f$, and trusted nodes, $t$, and a fixed Byzantine eviction rate.
We vary $f$ from 10\% to 30\% with a step of 2\%, $t$ from 1\% to 50\%, and the eviction rate from 0\% to 100\%.
We also consider the adaptive eviction rate mechanism which does not require setting a specific value.
While honest nodes follow the protocol, Byzantine nodes push faulty IDs to correct nodes and always return faulty IDs to pull requests.
We set the view size to 200 entries.





Figures~\ref{fig:0} to~\ref{fig:100} show the results of our experiments for  different configurations of the Byzantine eviction rate.
Their corresponding subfigures \textit{a}, \textit{b}, and \textit{c} represent resilience improvements, system-discovery overhead, and view-stability overhead as a function of the proportions of Byzantine and trusted nodes, respectively.
Finally, Figure~\ref{fig:adapt} shows the benefits of relying on our adaptive eviction rate mechanism.


\subsection{Resilience improvements}

As shown in Figures~\ref{fig:res0},~\ref{fig:res40}, and~\ref{fig:res60},
a minimum  of 1\% of trusted nodes ($t=1\%$) allows \protocol to decrease the proportion of Byzantine IDs in the views of honest nodes by 4\% under a 0\% eviction rate and by 7\% under a 60\% eviction rate
The incremental increase in $t$ up to 50\% provides further improvements in resilience, although sublinearly, up to 15\%.
Furthermore, the proportion $f$ of Byzantine nodes has no impact on resilience improvements.

Considering an eviction rate of 100\% might be a good idea at first glance.
Indeed, for as few as 10\% of Byzantine nodes, Figure~\ref{fig:res100} shows that this configuration outperforms the previous ones, providing resilience improvements ranging from 11\% to 21\%. 
However, as soon as $f$ increases, the improvements drop, reaching between 0 and 5\% percent with a Byzantine ratio of 30\%, highlighting the questionable aspect of increasing the eviction rate to such an extent.
A 100\% eviction rate delays the process by which trusted nodes learn about other trusted nodes since only pushed IDs from the \brahms protocol can enter their views.
This slowdown favors Byzantine nodes because they can pollute the views of honest nodes faster than trusted nodes could meet.

As shown in Figure~\ref{fig:resadapt}, the curves of the adaptive eviction rate have a similar shape to the 100\%-eviction-rate case.
However, the resilience improvements are better than any of the fixed rate scenarios described above when the fraction of Byzantine nodes is below 26\%.
With only 10\% of Byzantine nodes, \protocol offers resilience improvements of between 18\% and 25\% over \brahms.

\subsection{Performance overhead}

The overhead in terms of system discovery and view stability of \protocol increases when the eviction rate increases.
However, relatively high eviction rates provide better protection against view poisoning.
The adaptive eviction-rate configuration provides optimal results with overhead close to that of the 0\%-eviction-rate configuration as well as the highest level of resilience.
Increasing $t$ drives up the overhead to a tipping point when larger shares of trusted communication compensate for the propagation of Byzantine IDs that slow down the discovery of non-Byzantine nodes.
The overhead values of system discovery and view stability are closely related, and we can draw similar trends and interpretations.
For this reason, we only comment on the system-discovery overhead in the remainder of this section. 

Figure~\ref{fig:couv0} shows the system-discovery overhead of \protocol with a 0\% eviction rate.
This overhead is negligible and does not exceed 3\% when the number of trusted nodes is very small ($t=1\%$).
Indeed, in this configuration, trusted nodes can hardly find their siblings in the mass, making trusted communication rare and making it unlikely that malicious nodes will send Byzantine IDs to trusted nodes.
Moreover, this overhead remains constant regardless of the proportion of Byzantine nodes. 
In other words, since trusted nodes do not evict any ID from pull responses, discovery time is barely affected by Byzantine attempts to pollute views compared with our baseline results with \brahms.

The system-discovery time behaves differently as the share of trusted nodes increases.
When $f$ is less than 14-16\%, the time required for system discovery is actually lower than in the case of \brahms, with a maximum gain of 18\% when considering 50\% of trusted nodes.
With an eviction rate of 0\%, trusted nodes can quickly discover their siblings, leaving the field open for trusted communication to take over with efficient dissemination of IDs, sending them back to the honest nodes that run \brahms.
Nonetheless, when $f$ exceeds 14-16\%, attacks by Byzantine nodes to poison the views of honest nodes hinder the benefits of efficient trusted communication, and the protocol suffers an overhead of 3-16\% depending on the shares of trusted nodes considered.

A particular pattern of discovery time should be noted.
For $f$ greater than or equal to 16\%, the most extreme values of $t$ lead to minimal overhead.
Conversely, intermediate values (\ie $t=10$ and $t=20$) incur the highest overhead.
This bell-shaped pattern results from  trusted nodes' having Byzantine identifiers in their views that are advertised by malicious nodes and their sharing them with other trusted nodes.
As a result, this rapid dissemination of Byzantine identifiers slows the discovery time of the entire system, which relatively small shares of trusted nodes (\ie small values of $t$) cannot compensate for.
A tipping point is reached once $t$ exceeds 20\%.
Overhead decreases, resulting in a higher proportion of trusted nodes broadcasting more non-Byzantine identifiers.

A similar but amplified bell curve is observed in Figures~\ref{fig:couv40} and~\ref{fig:couv60} where the eviction rate reaches 40\% and 60\% respectively.
This exacerbated effect results from the fact that the total number of evicted identifiers increases as $t$ increases from 1 to 10-20\%, resulting in an additional delay in the discovery of the overall system which again is not compensated for by the relatively small share of trusted nodes.

Another trend is visible on all eviction-rate configurations, except for 0\% (Figures~\ref{fig:couv40},~\ref{fig:couv60},~\ref{fig:couv100},~\ref{fig:couvadapt}).
Overhead costs decrease as the proportion of Byzantines increases from less to more than 18-20\%.
Indeed, increasing $f$ guides both the amount of non-Byzantine nodes and the amount of over-advertised IDs to non-Byzantine nodes.
Therefore, fewer non-Byzantine identifiers and more over-advertised IDs have to be discovered,  resulting in faster system discovery compared to lower values of $f$.

The adaptive eviction-rate mechanism provides an optimal trade-off between performance overhead and resilience.
With $t$ as small as 1\%, resilience is improved by 8-18\%; system discovery and view stability require only 15-20\% more rounds.

\section{Security analysis}
\label{sec:sec}

This section provides a security analysis of \protocol against two attack vectors: trusted node identification, and view-poisoned trusted node injection.

\subsection{Trusted node re-identification}

In the original \brahms paper~\cite{bortnikov2009brahms}, the authors
prove that an adversary that tries to isolate nodes by launching
targeted attacks on them cannot partition the network.
However, \protocol's use of trusted communications makes trusted nodes a particularly attractive target for the adversary.
If the adversary could identify these nodes, it could launch a targeted attack  to pollute their views with Byzantine IDs.
The adversary would thus have a backdoor to inject Byzantine IDs into the protocol to increase the dissemination speed of Byzantine IDs.
This could inhibit the benefits of trusted communications on resilience to Byzantine behaviors and could further increase the time required to reach stability and to discover new peers.
To evaluate this attack feasibility, we must distinguish between two cases: after reaching view stability and before.

We recall that view stability is defined as the state in which the maximum difference in view composition between any non-Byzantine node and the average of all non-Byzantine nodes does not exceed 10\%.
Our experiments show that once this state is reached, the difference in view composition between trusted and untrusted nodes does not exceed 2\% on average.
This difference is therefore much smaller than the 10\% threshold used to define view stability.
This makes it very difficult for an adversary to identify trusted nodes even with global knowledge of all views in the system.


On the other hand, before reaching view stability, trusted nodes propagate their views to untrusted nodes via answers to pull requests, which are sometimes (and purposely) significantly different from the views of untrusted nodes. 
To assess the extent to which this can lead to successful identification, we consider an attack in which each Byzantine node measures the proportion of Byzantine IDs in the pull answers it receives from each non-Byzantine node and provides this information to the adversary.
The adversary first computes the average percentage of Byzantine IDs in
the views of all honest nodes.
Then, for each honest node, it computes the
difference between this average percentage and the percentage of
Byzantine IDs in the honest node's view.
If the difference exceeds the threshold, the adversary labels the node as a trusted node.
We experimentally tested several thresholds and the one that
maximizes the identification outcome is 10\%.

\begin{figure*}[!htb]
  \centering     
  \subfloat[Recall]{
    \label{fig:recall10}\includegraphics[width=0.32\linewidth]{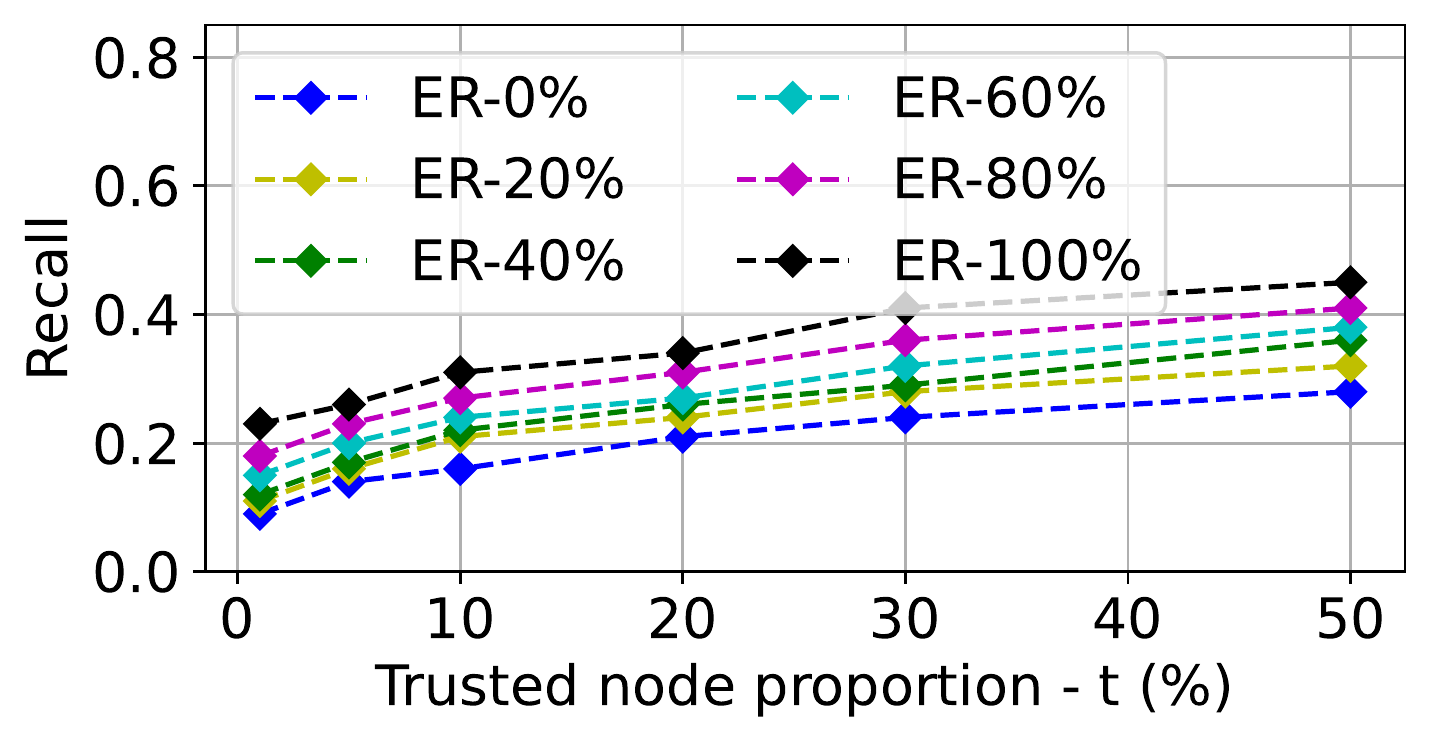}
  }
  \subfloat[Precision]{
    \label{fig:prec10}\includegraphics[width=0.32\linewidth]{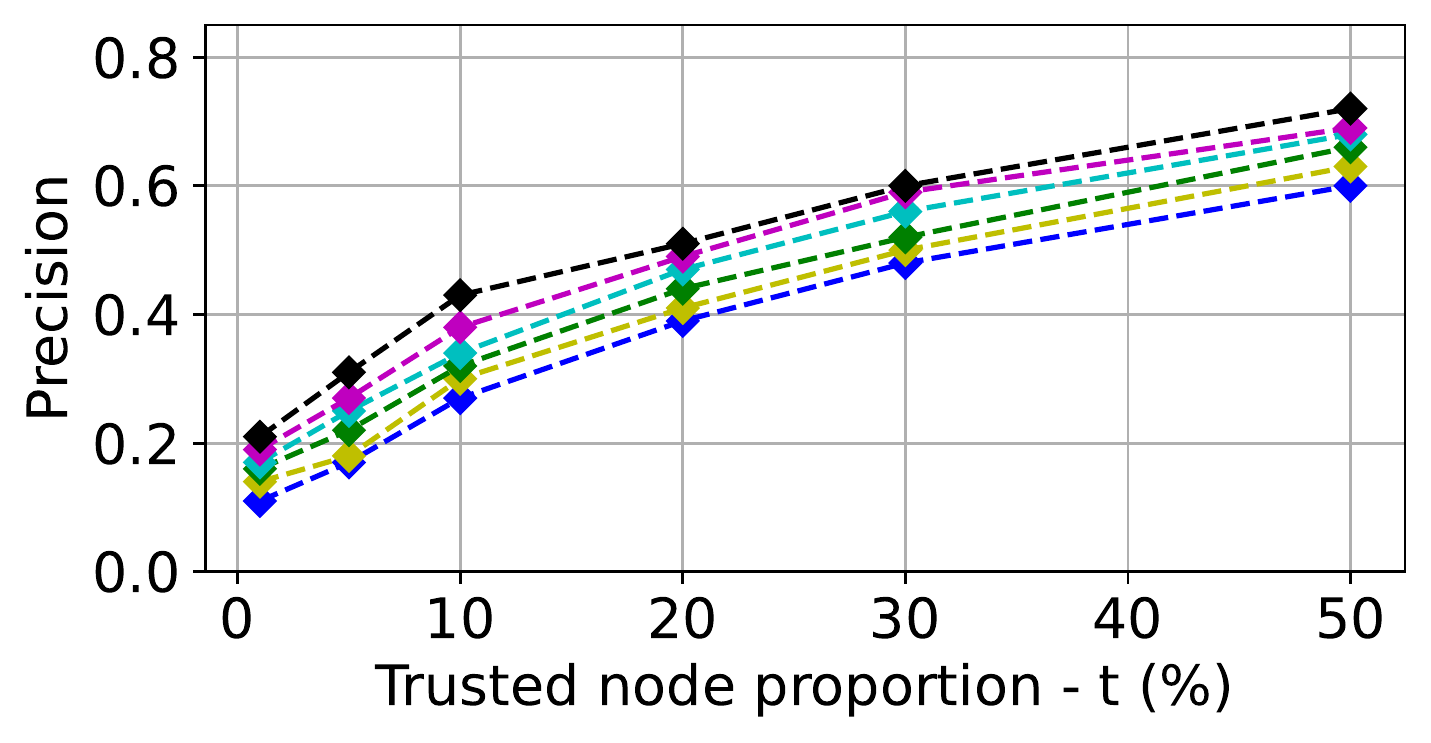}
  }
  \subfloat[F1-score]{
    \label{fig:f10}\includegraphics[width=0.32\linewidth]{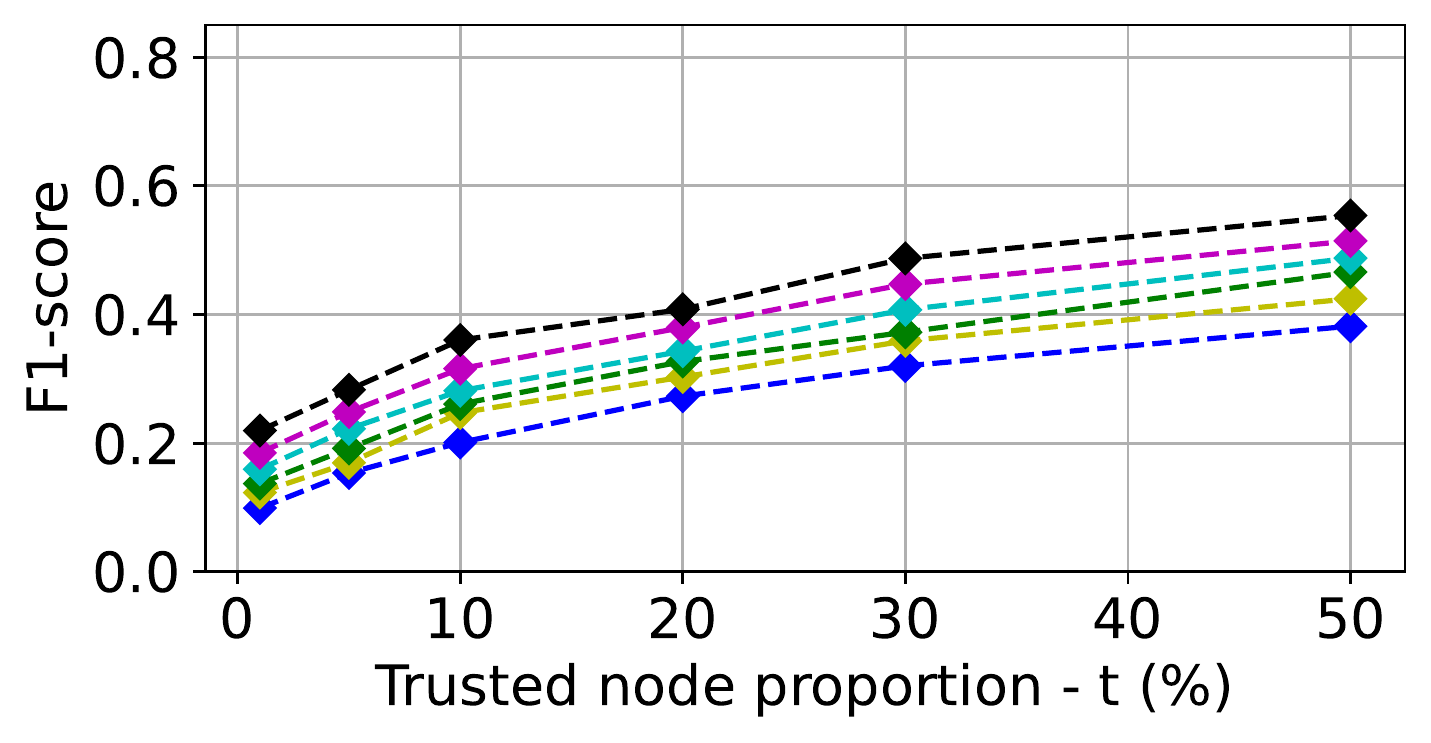}
  }
  \caption{Precision, recall and F1-score of trusted-node identification under 10\% of Byzantine nodes\vspace{-0.1cm}}
  \label{fig:ident10}

  \centering     
  \subfloat[Recall]{
    \label{fig:recall30}\includegraphics[width=0.32\linewidth]{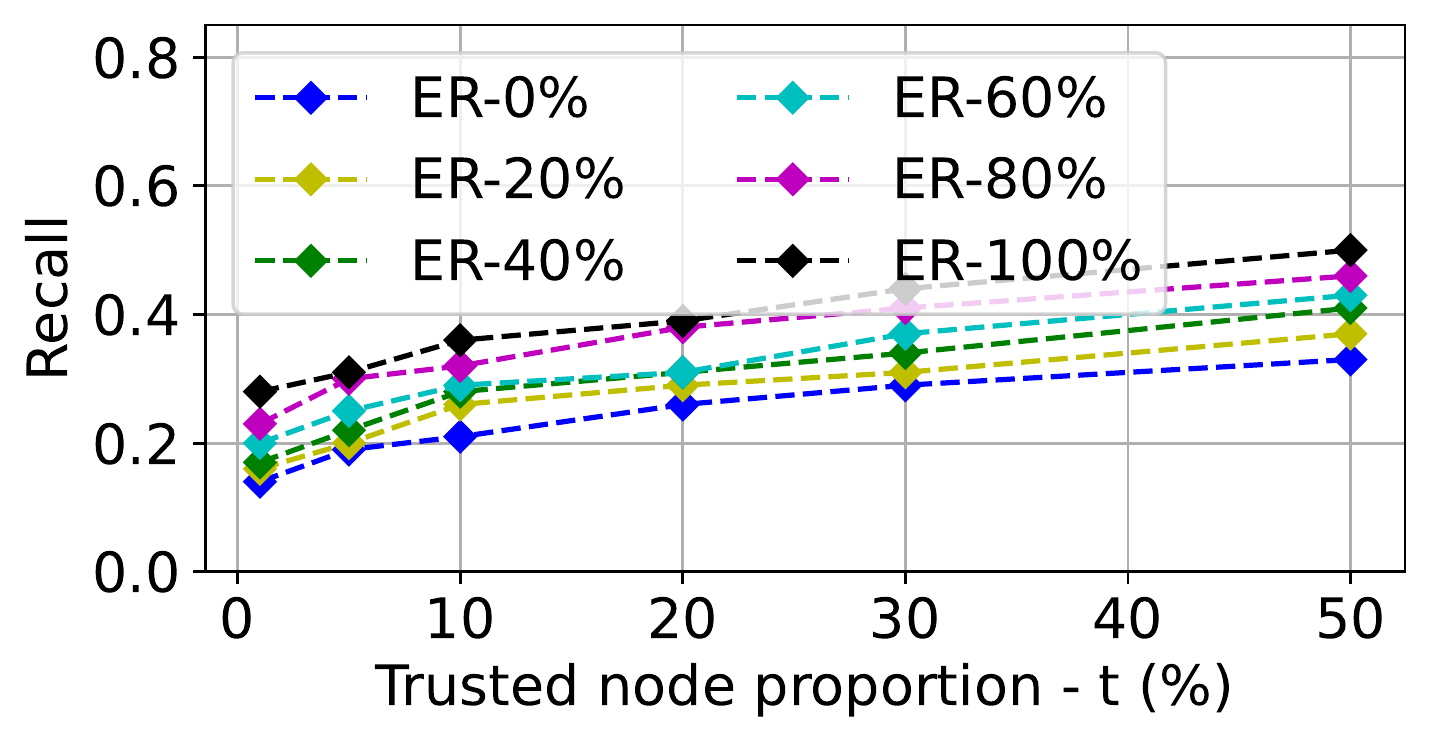}
    }
  \subfloat[Precision]{
    \label{fig:prec30}\includegraphics[width=0.32\linewidth]{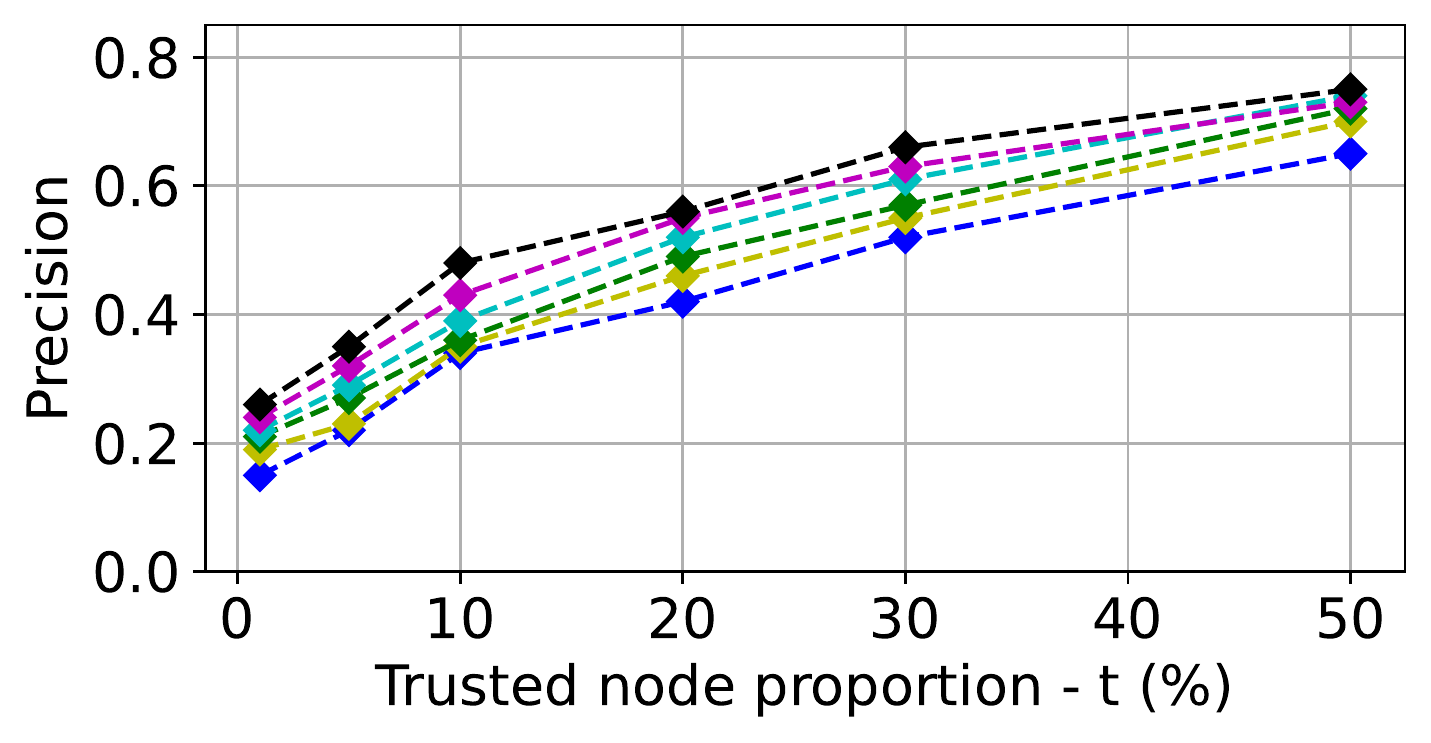}
  }
  \subfloat[F1-score]{
    \label{fig:f30}\includegraphics[width=0.32\linewidth]{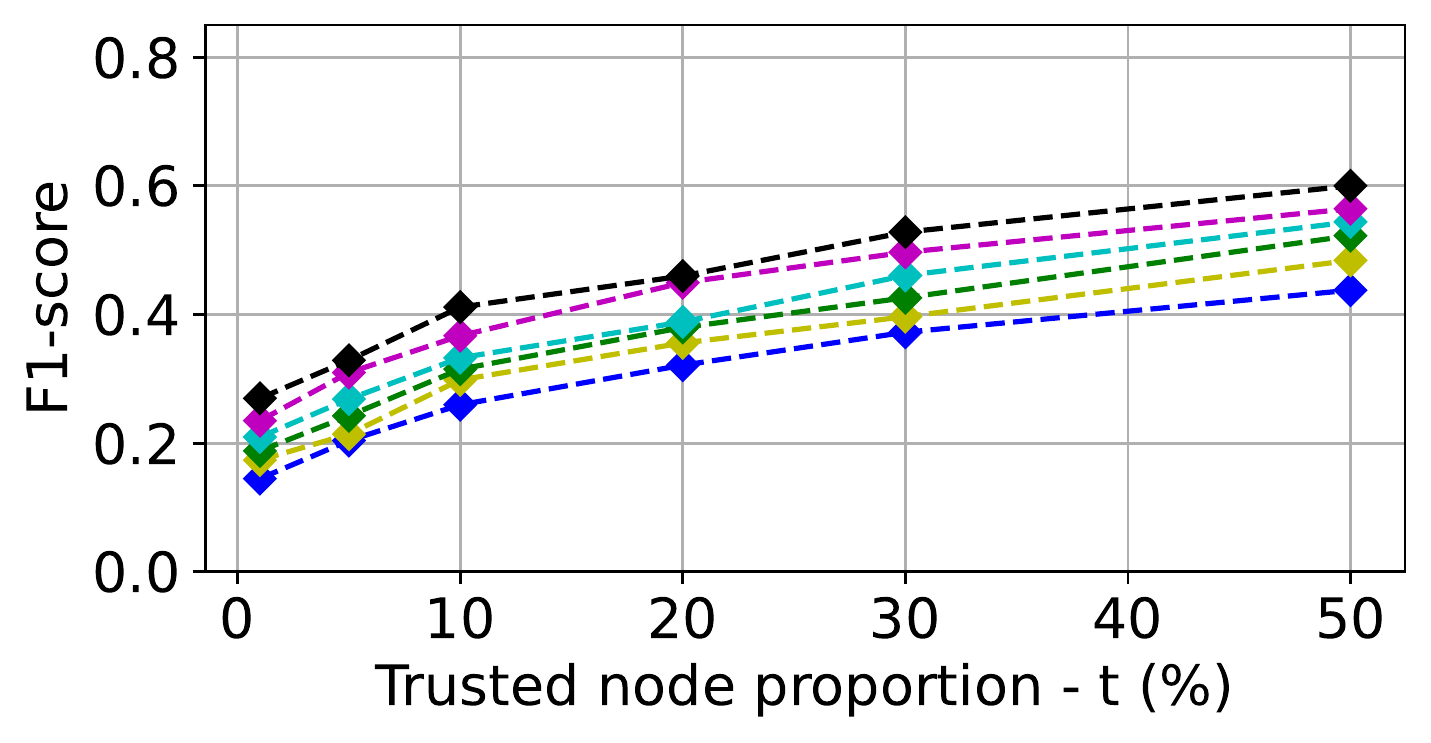}
  }
  \caption{Precision, recall and F1-score of trusted-node identification under 30\% of Byzantine nodes \vspace{-0.1cm}}
  \label{fig:ident30}
  \subfloat[Identificaiton recall]{
    \label{fig:recadapt10}\includegraphics[width=0.32\textwidth]{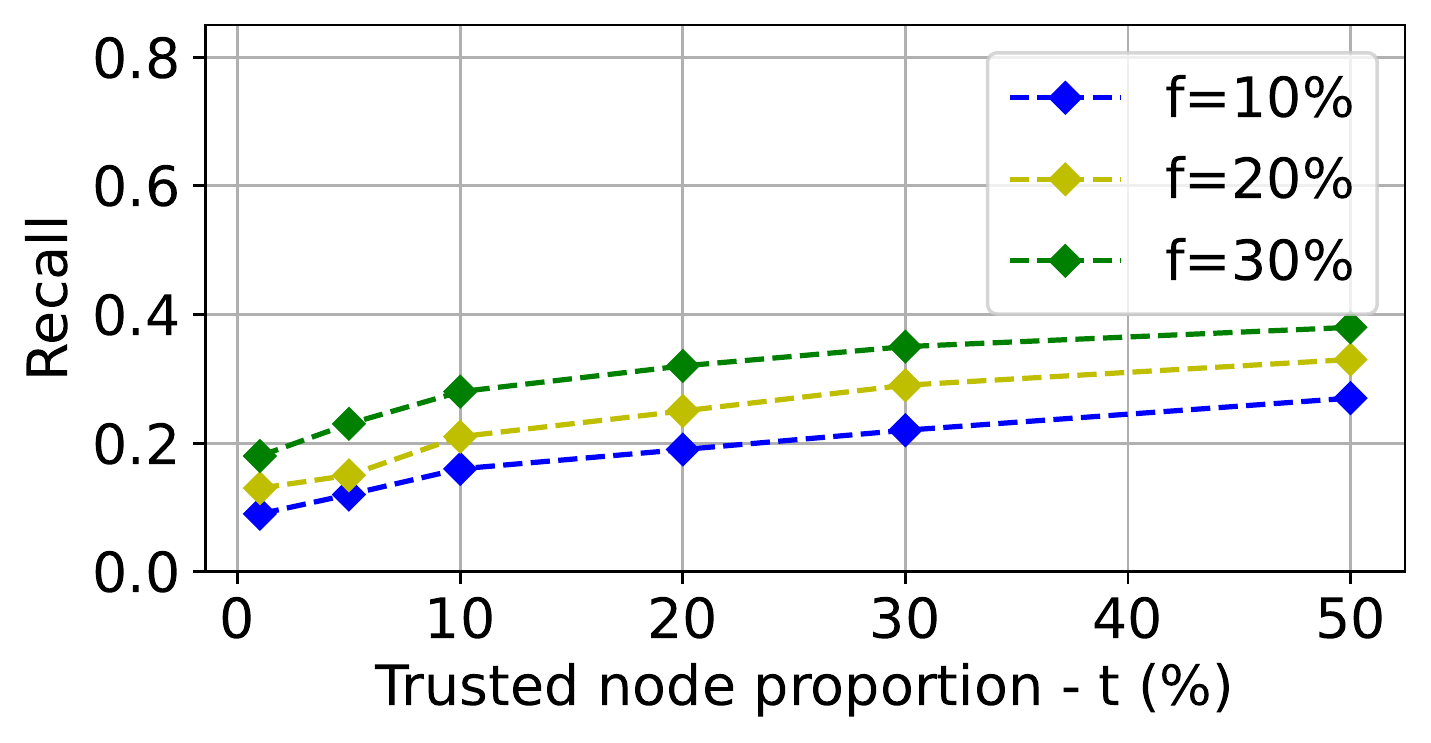}
    }
  \subfloat[Identification precision]{
    \label{fig:precadapt10}\includegraphics[width=0.32\textwidth]{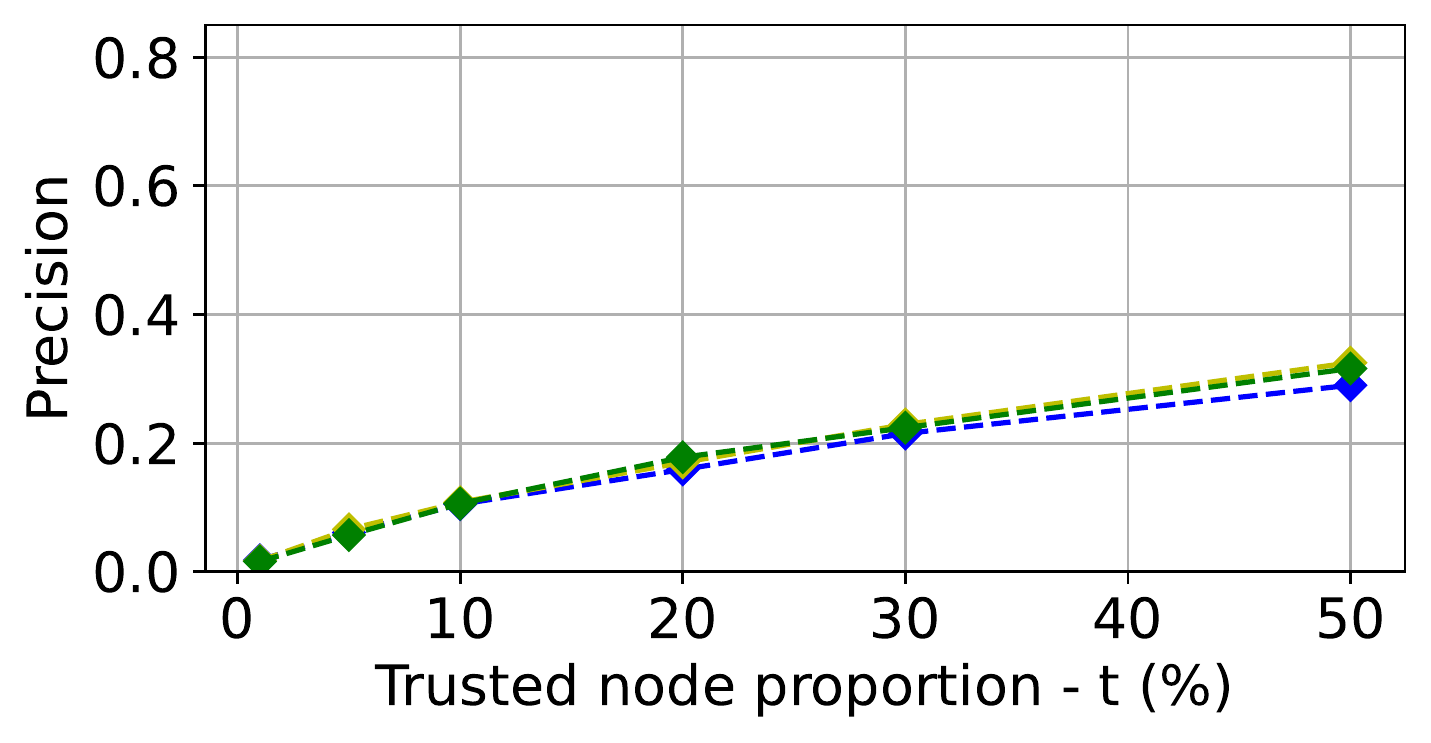}
  }
  \subfloat[Identification F1-score]{
    \label{fig:f1scoreadapt10}\includegraphics[width=0.32\textwidth]{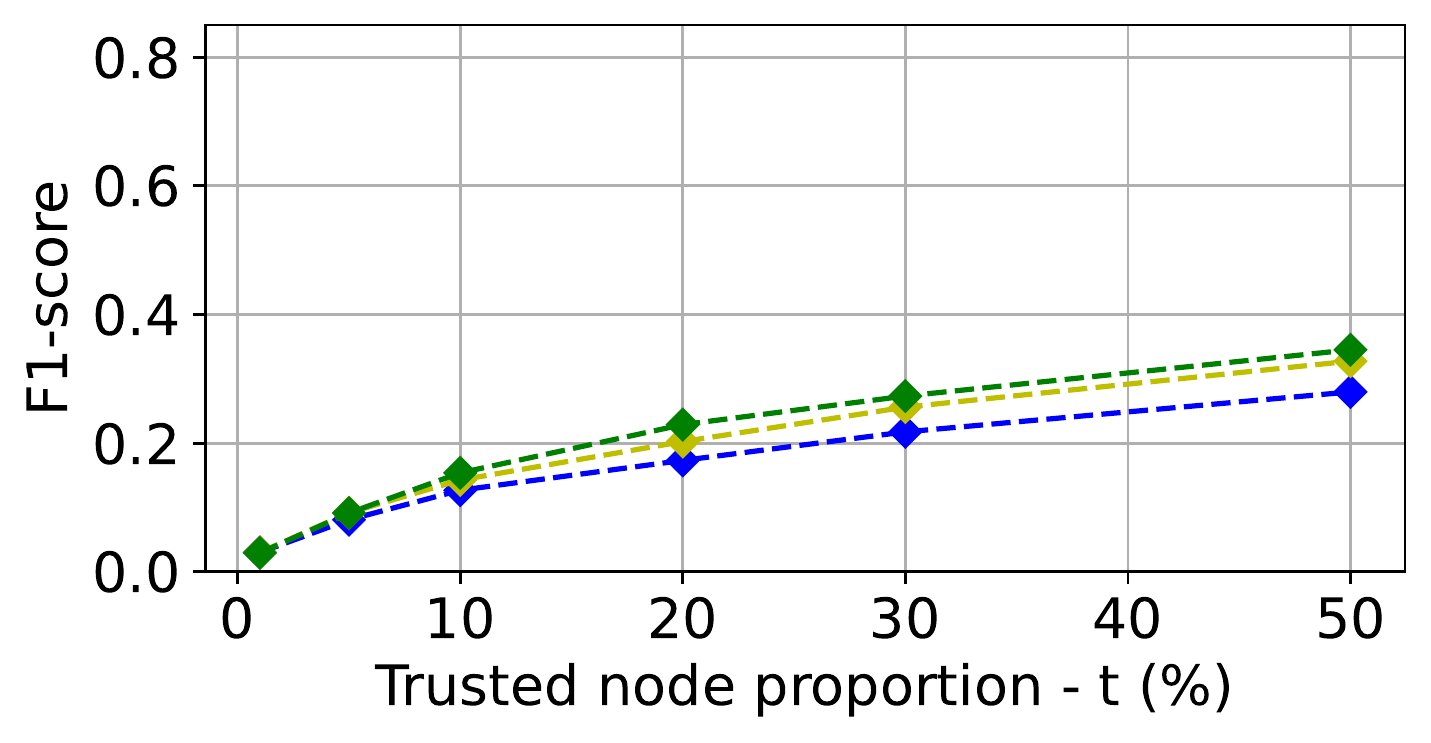}
    }
    \caption{Precision, recall and F1-score of trusted-node identification under adaptive eviction rate\vspace{-0.1cm}}
    \label{fig:identadapt}

  \centering     
  \subfloat[Attack on a system with $t=1\%$]{
    \label{fig:sgx_attack_1_2}\includegraphics[width=0.32\linewidth]{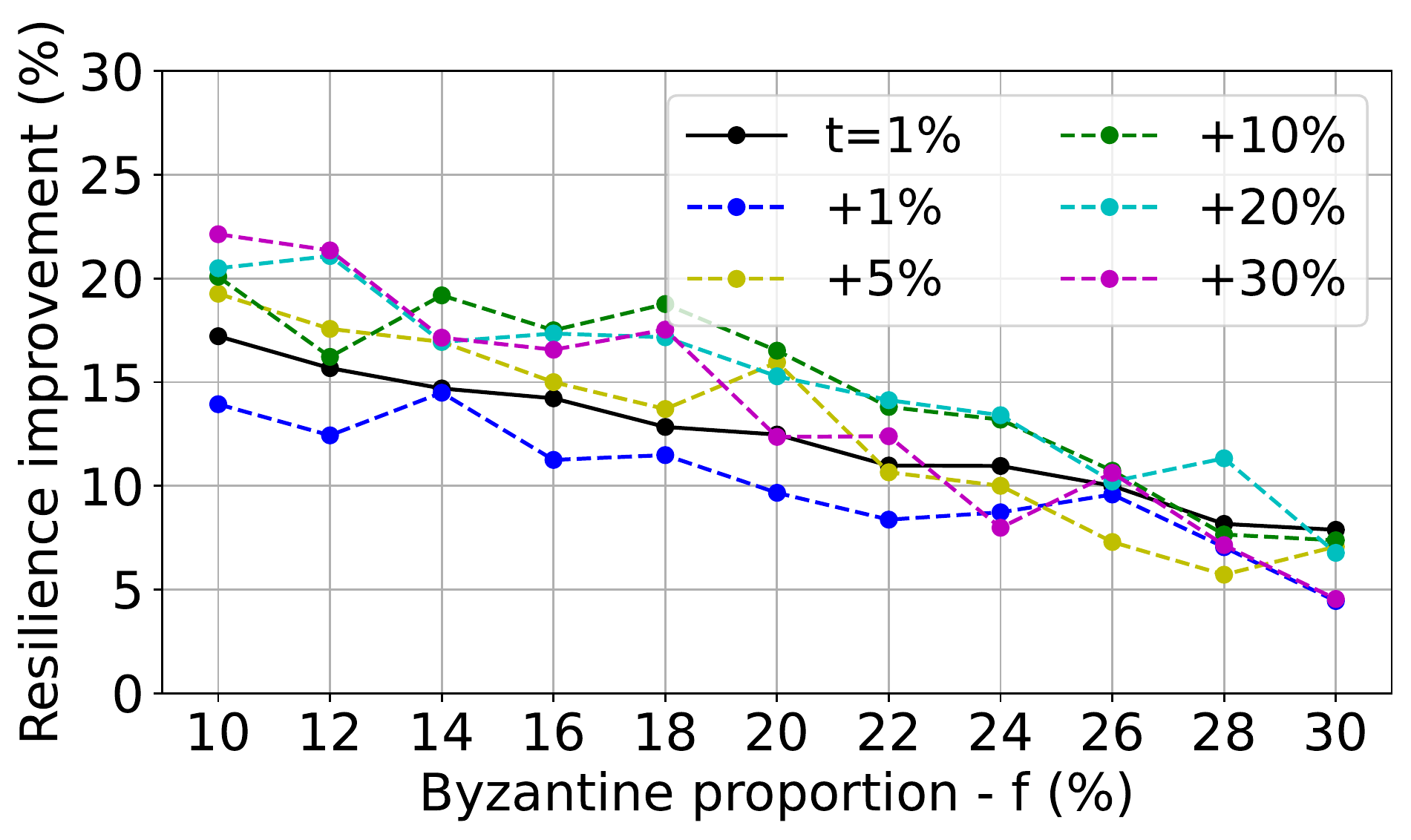}
  }
  \subfloat[Attack on a system with $t=10\%$]{
    \label{fig:sgx_attack_10_2}\includegraphics[width=0.32\linewidth]{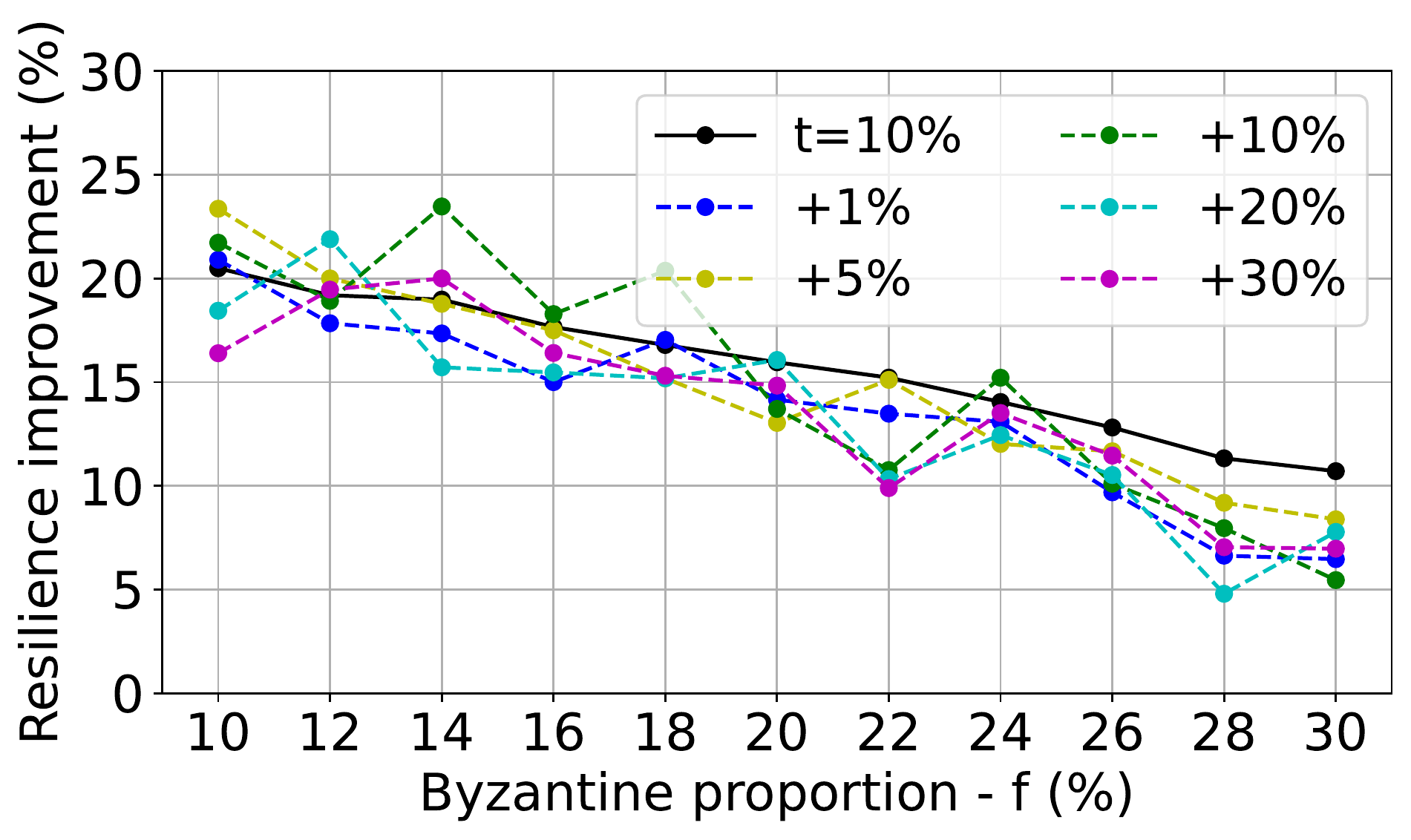}
  }
  \subfloat[Attack on a system with $t=30\%$]{
    \label{fig:sgx_attack_30_2}\includegraphics[width=0.32\linewidth]{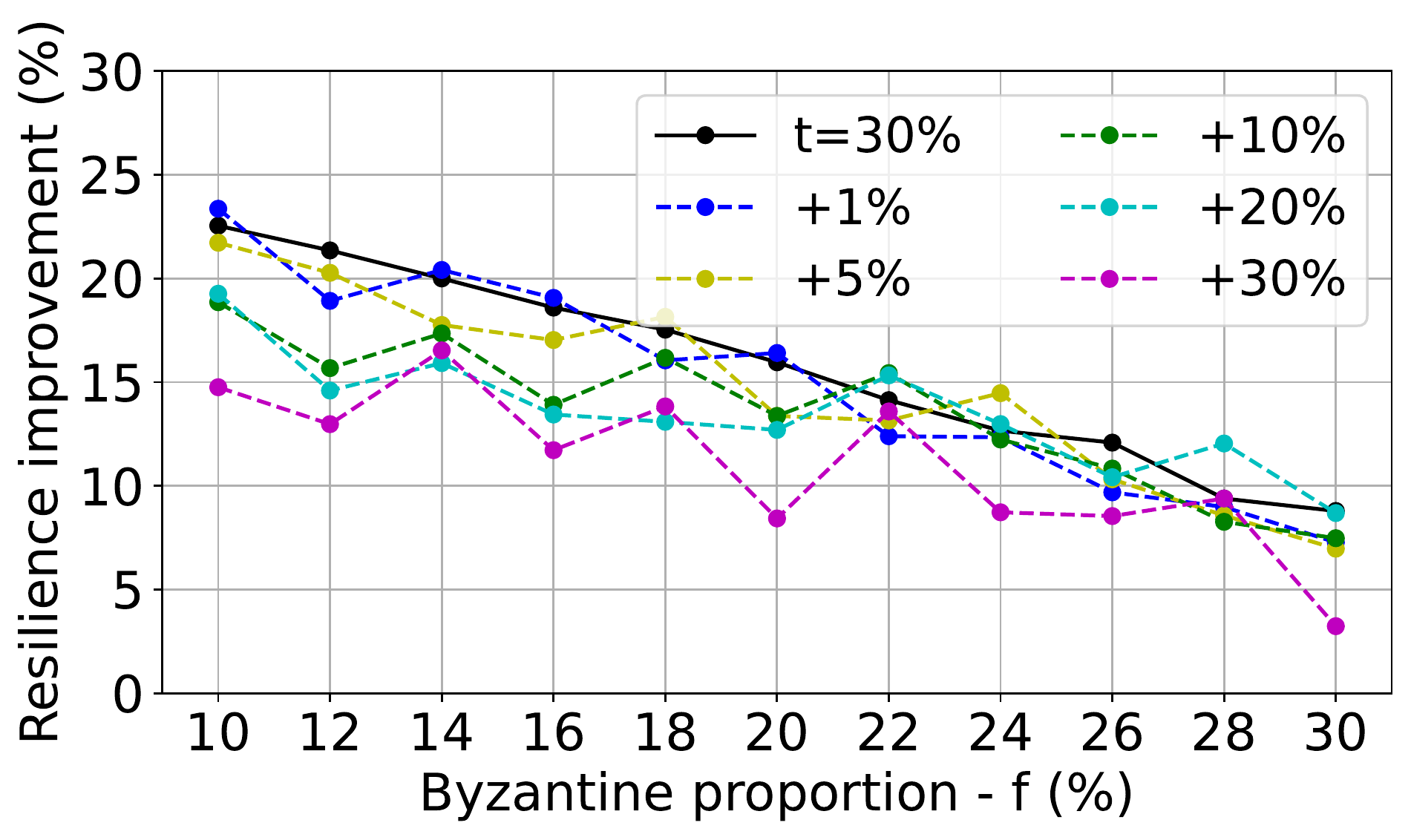}
  }
  \caption{Corrupted trusted node injection\vspace{-0.1cm}}
  \label{fig:attack_SGX}

\end{figure*}

We evaluate the effectiveness of this attack with $10\%$ of Byzantine
nodes in Figures~\ref{fig:ident10} and $30\%$ of Byzantine nodes in
Figure~\ref{fig:ident30}.  For each Byzantine configuration, we
present recall (subfigures \ref{sub@fig:recall10}),
precision (subfigures \ref{sub@fig:prec10}), and F1-score (subfigures
\ref{sub@fig:f10}) values for different values of eviction-rate (denoted $ER$ in
the figures) and shares of trusted node. Figure~\ref{fig:identadapt}
highlights the results with an adaptive eviction rate.

Results show that the effectiveness of the attack grows steadily with
the percentage of trusted nodes in the system.
It is clear that the more nodes of trust there are, the easier it is for the adversary to identify some of them.
However, success rate also depends on the eviction rate.
In particular, Figure~\ref{fig:ident10} and Figure~\ref{fig:ident30} show an increase in recall of $0.14$ ($14\%$) when the eviction rate is increased from $0$ to $100\%$, and an increase in precision of about $0.12$ under the same conditions.
An eviction rate of $60\%$, for example, leads to an identification precision of $50\%$ and a recall of about $30\%$  with $20\%$ of trusted nodes. Moreover, these values increase only slightly with the percentage of Byzantine nodes.

Although the above values may seem high, they are drastically reduced by adopting an adaptive eviction rate. Figure~\ref{fig:identadapt} shows that precision varies between $0$ and $0.3$ when going from $1$ to $50\%$ of trusted nodes and does not significantly depend on the proportion of Byzantine nodes. Recall follows a similar pattern, although with a stronger dependence on the percentage of Byzantine nodes.  Its values range from $0.01$ with $10\%$ of Byzantine nodes, and $0.30$ with $30\%$ of Byzantine nodes.

\subsection{View-poisoned trusted node injection}

We have shown that trying to identify trusted nodes in order to isolate them
bring only limited results for the adversary, especially with an
adaptive eviction rate.
As an alternative, the adversary can adopt a totally different strategy by purchasing SGX-capable devices and using them as a means of disseminating Byzantine node identifiers to real trusted nodes.

Specifically, the adversary can deploy some SGX-capable nodes in a network that contains only Byzantine nodes, in order to fill their views with Byzantine identifiers.
The adversary can then move these view-poisoned trusted nodes into the actual network and wait for them to disseminate the Byzantine identifiers to the actual trusted nodes.

Figure~\ref{fig:attack_SGX} shows the effect of this attack in a network with different initial proportions of trusted nodes, $t=1\%$, $10\%$, and $30\%$.
Each plot shows how the resilience-improvement metric varies as a function of the proportion of Byzantine nodes with several percentages of view-poisoned trusted nodes added by the adversary $1\%$, $5\%$, $10\%$, $20\%$, and $30\%$.
In all the plots, the black line represents the baseline with $t$ honest trusted nodes and no attacks.

When the proportion of honest trusted nodes is small, $t=1\%$, and for a relatively small proportion of Byzantine nodes, the addition of view-poisoned trusted nodes does not appear to significantly harm resilience.
Figure~\ref{fig:sgx_attack_1_2} shows that adding a large number of view-poisoned trusted nodes even seems to improve resilience.
Indeed, even view-poisoned nodes are forced to work with correct \brahms implementations.
As a result, all trusted nodes, including view-poisoned ones, end up removing the overrepresented identifiers, and the view-poisoned  nodes end up reinforcing the trusted portion of the network.
As the proportion of Byzantine nodes increases (to the
right of the plot in Figure~\ref{fig:sgx_attack_1_2}), the sampling
process becomes less effective at removing Byzantine identifiers from
the views, thus decreasing the percentage resilience improvement.

As the  proportion of honest trusted nodes increases from $t=10\%$ in Figure~\ref{fig:sgx_attack_10_2} to $t=30\%$ in Figure~\ref{fig:sgx_attack_30_2}, the benefit of having additional trusted nodes with a small portion of Byzantine nodes disappears as adding a fraction of trusted nodes to a larger set makes less of a difference and instead becomes significantly counterproductive, when $t=30\%$. 

\section{Discussion}
\label{sec:discussion}
Our experimental evaluation in Section~\ref{sec:exp} shows that
\protocol can provide significant resilience improvement over
\brahms. Its adaptive eviction strategy provides a $17\%$ resilience
improvement in the presence of $10\%$ of Byzantine nodes at the
minimal cost of $1\%$ of trusted nodes. Even when the proportion
of Byzantine nodes increases to $30\%$, \protocol still provides an
$8\%$ improvement.

Most importantly, our analysis in Section~\ref{sec:sec} also shows
that \protocol with $1\%$ of trusted nodes effectively resists attackers that can (i) try to identify the trusted nodes in order to
isolate them, or (ii) inject view-poisoned trusted nodes that try to
spread malicious identifiers to honest nodes. With respect to (i), an
attacker can identify only less than $10\%$ of the trusted nodes with
precision below $2\%$ for an F1-score of less than $3\%$.  When it
is possible to increase the proportion of honest trusted nodes to
$10\%$, \protocol provides even higher levels of resilience
improvement above $20\%$, with trusted nodes becoming only slightly
more detectable (precision and F1-score below $15\%$).  With respect
to (ii), an attack with $5\%$ of view-poisoned trusted nodes in a
system with $1\%$ of honest trusted nodes even improves resilience to
Byzantine nodes when the latter's proportion is not too high (below
$20\%$).  Together these results suggest that \protocol can serve as
an effective overlay for a number of Byzantine-sensitive applications
like smart-contract platforms and cryptocurrencies.


\section{Related work}
\label{sec:rw}

Random peer sampling (RPS) has mostly been studied in non-adversarial
environments. Cyclon~\cite{voulgarisCYCLON2005},
Newscast~\cite{tolgyesi2009adaptive}, and the more generic protocol
framework for Gossip based peer sampling of Jelasity \emph{et
al.}~\cite{jelasityGossipbased2007} efficiently handle churn and quickly
discover other nodes in the network. Our contribution uses such a RPS
protocol between trusted nodes. However, these protocols are easily disturbed
by Byzantine nodes performing eclipse or poisoning attacks.

A few studies targeted the problem of making RPS tolerant to Byzantine
nodes~\cite{jesi2010secure,bortnikov2009brahms,anceaume2013uniform,anceaume2013power}. In
the \emph{Secure peer Sampling} framework~\cite{jesi2010secure}, each node uses a detection
mechanism to identify and blacklist maliciously acting nodes. This
protocol remains, however, vulnerable to rapid flooding attack as correct nodes cannot
identify and blacklist attackers before being overwhelmed by them and
isolated. \protocol reduces for trusted nodes the risk of isolation through such attacks by dropping pull requests answers from untrusted nodes.


Instead of using a detection mechanism, Brahms~\cite{bortnikov2009brahms}
employs push limiting and view sampling in order to mitigate the
over-representation of malicious nodes in the view of honest ones. The use
of a Trusted Execution Environment in \protocol enables better performance
than Brahms in terms of resilience against malicious nodes and alleviates
some of the significant running time overhead of Brahms.

Anceaume \emph{et al.}~\cite{anceaume:inria-00617866} formally analyze the
requirements of peer sampling protocols when these need to resist malicious
attacks. They observe that a necessary condition lies in limiting the
request rates of nodes. \protocol  complements this measure with its eviction-rate policy.
The same
authors~\cite{anceaume2013uniform,anceaume2013power}
employ count-min sketches to unbias a biased stream of
identifiers. Adopting a similar technique in \protocol could
constitute  interesting  future work.

With respect to resistance to Sybil attacks, HAPS~\cite{amaury}
uses a prefix tree to limit the number of known
identifiers in each IP-address prefix. This effectively limits Sybil
attacks under the assumption that honest nodes are uniformly
distributed over the IP address space. This idea is orthogonal to our
rate-limiting mechanism and it could constitute an interesting
addition to \protocol a large-scale deployment.

Finally, GossipSub~\cite{vyzovitis2020gossipsub} is a recent proposal for strengthening the gossip-based dissemination in the Ethereum 2.0 and Filecoin networks.
These open blockchain systems rely on gossip to quickly propagate of transactions and blocks between miners.
GossipSub implements an ad hoc mesh construction protocol that maintains a balanced in- and out degree for nodes, as do protocols from the peer sampling family. 
This protocol does not, however, target diversity, i.e., the constant renewal of peers in the views of each node.
This  makes it specific for dissemination and unfit for other classes of gossip-based protocols such as overlay construction~\cite{jelasity2009t,voulgarisVICINITY2013} or self-organizing aggregation~\cite{jelasity2005gossip}.
GossipSub reduces the risk of attacks by implementing a reputation mechanism.
Every node ranks peers in its view, and peers are incentivized to follow the protocol for a long period of time and with the same communication partners.


\section{Conclusion}
\label{sec:conclusion}

We presented \protocol, a novel Byzantine-tolerant random peer sampling protocol that builds and leverages trusted gossip-based communication.
\protocol interoperates with \brahms, the most resilient peer sampling protocol to date, 
and reduces the impact of a poisoning attack against its nodes.
Our experiments show that with only 1\% of SGX-capable devices, \protocol can reduce the proportion of Byzantine IDs in the views of honest nodes by up to 17\% when the system contains 10\% of  Byzantine nodes, at the cost of very limited overhead.
In addition, the security guarantees of \protocol hold even in the presence of a powerful attacker attempting to identify trusted nodes and injecting corrupted trusted nodes.


\balance
\bibliographystyle{IEEEtran}
\bibliography{biblio_rpsgx.bib}

\end{document}